\documentclass[a4paper, conference]{IEEEtran}
\IEEEoverridecommandlockouts
\usepackage{cite}
\usepackage{amsmath,amssymb,amsfonts}

\usepackage{graphicx}
\usepackage{algorithmic}
\usepackage{algorithm}
\usepackage{graphicx}
\usepackage{textcomp}
\usepackage{xcolor}
\usepackage{cleveref}
\usepackage{booktabs}
\usepackage{paralist}
\usepackage{subfigure}
\usepackage{multirow} 
\def\BibTeX{{\rm B\kern-.05em{\sc i\kern-.025em b}\kern-.08em
    T\kern-.1667em\lower.7ex\hbox{E}\kern-.125emX}}

\begin{document}

\title{Collaborative Resource Management and Workloads Scheduling
in Cloud-Assisted Mobile Edge Computing across Timescales
\thanks{*Corresponding authors.}
}

\author{
{Lujie Tang$^{1,2}$, Minxian Xu$^{1,*}$, Chengzhong Xu$^3$ and  Kejiang Ye$^{1,*}$}
\vspace{1.6mm}\\
\fontsize{12}{10}\selectfont\itshape
\,$^{1}$    Shenzhen Institute of Advanced Technology, Chinese Academy of Sciences\\
\,$^{2}$    University of Chinese Academy of Sciences \\
\,$^3$ University of Macau \\
\,\{lj.tang, mx.xu, kj.ye\}@siat.ac.cn, czxu@um.edu.mo\\
}

\maketitle

\begin{abstract}
Due to the limited resource capacity of edge servers and the high purchase costs of edge resources,  service providers are facing the new challenge of how to take full advantage of the constrained edge resources for Internet of Things (IoT) service hosting and task scheduling to maximize system performance. In this paper, we study the joint optimization problem on service placement, resource provisioning, and workloads scheduling under resource and budget constraints, which  is formulated as a mixed integer non-linear programming problem.
Given that the frequent service placement and resource provisioning will significantly increase system configuration costs and instability,
we propose
a two-timescale framework for  
resource management and workloads scheduling, named RMWS.
RMWS consists of a Gibbs sampling algorithm
and an alternating minimization algorithm
to determine the service placement
and resource provisioning 
on large timescales. 
And a sub-gradient descent method has been designed to solve the workload scheduling challenge on small timescales.
We conduct comprehensive experiments
under different parameter settings. The RMWS consistently ensures a minimum 10\% performance enhancement compared to other algorithms, showcasing its superiority. 
Theoretical proofs are also provided accordingly.
\end{abstract}

\begin{IEEEkeywords}
Mobile Edge Computing, Service placement, Resource provisioning, Workloads scheduling, Across Timescales
\end{IEEEkeywords}

\section{Introduction}
In recent decades, 
the rapid development of IoT devices
and  network
communication technologies have significantly
contributed to the proliferation
of  IoT application services\cite{vimal2020enhanced,mukhopadhyay2021artificial,cui2021integrating}.
The emergence of IoT application services,
such as smart transportation, smart factory\cite{siqueira2021service}, smart city\cite{xu2020intelligent}, and
smart home, etc,
have captured the attention 
and interest of the general public.
% In recent decades, 
% the swift evolution of IoT devices and network communication technologies have significantly
% contributed to the proliferation
% of  IoT application services\cite{vimal2020enhanced,mukhopadhyay2021artificial,cui2021integrating}. This emergence has garnered widespread attention and interest from the general public\cite{siqueira2021service,xu2020intelligent}.
Typically, IoT devices are constrained
by physical dimensions and battery life,
making it challenging to meet
the computational resource
and capability requirements of
different application services\cite{goudarzi2022scheduling}.
Cloud computing is a key solution to the resource constraints of IoT devices.
However, the wide area network
connecting traditional cloud servers
to IoT devices can
introduce unpredictable communication delays
and jitter, potentially causing 
adverse effects
on latency-sensitive application services.

As an emerging promising computing paradigm,
Mobile Edge Computing (MEC)
has attracted growing attention 
from both industry and academia~\cite{wang2023edge,brecko2022federated,du2023computation}.
The distributed architecture of MEC
enables the deployment of services
and data closer to end-users,
leading to significant reductions
in response latency and data transmission costs.
The appeal of these advantages
has prompted more service providers
to transition their operations
to MEC platforms 
to meet the growing demand
for real-time services.
However, compared to the
cloud with elastic resource capacity,
the resources in edge nodes are limited.
Only a small number of services
or applications can be accommodated on edge servers. 

A more effective approach
to address resource scarcity is
to jointly utilize the computation,
storage and communication resources
of the nearby edge servers
and the remote cloud server,
maximizing edge-edge
and edge-cloud cooperation,
achieving load balancing
and minimizing execution latency.
Taking the cloud-assisted edge computing system architecture depicted in Figure 1 as
an example, each edge node has its own region coverage restrictions and resource capacity due to hardware constraints.
IoT devices in different regions will send service requests to the nearest edge node. If a device is unable to request a specified service,  the request will be routed to neighboring edge servers capable of handling the service request or to the cloud server for processing.
In Figure 1, it can be observed that different regions  have different types and numbers of workload requests. This also requires us to dynamically adjust service placement and resource provisioning based on actual workload conditions to ensure service performance. In this regard, service providers need to focus on
the following critical issues:
\textbf{service placement},
\textbf{resource provisioning}
and \textbf{workloads scheduling}.
In general,
service placement requires
dynamically optimizing the placement
of services on edge servers
to better utilize edge server resources~\cite{goudarzi2020application,long2022mobility,ray2023adaptive,xu2022pdma}.
Resource provisioning requires
the flexibility to adjust
resource provisioning
for each service to optimize
overall system performance\cite{li2022joint,smolka2023edgedecap,plachy2021dynamic}.
Additionally,
to balance system workloads
and improve system performance,
workloads scheduling ensures that
task requests are dynamically dispatched
to appropriate edge nodes
or remote cloud servers~\cite{ma2021dynamic,zhu2023smart,zhang2024incentive}.

\begin{figure}[tbp]
\centerline{\includegraphics[width=0.95\linewidth]{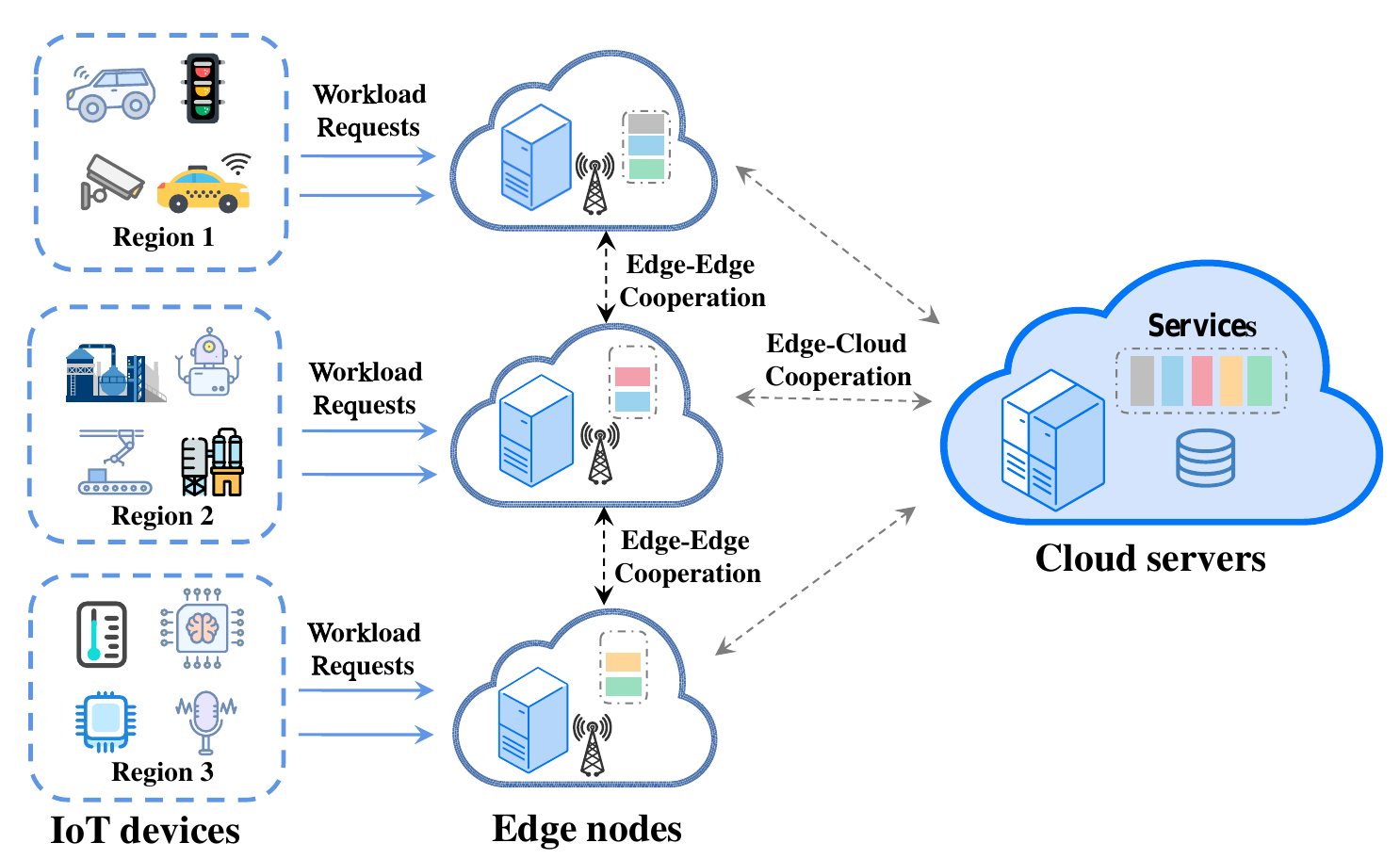}}
% \vspace{-8pt}
\caption{System model.}
\label{fig1}
\vspace{-10pt}
\end{figure}

However, the aforementioned
joint optimization problem
may encounter some challenges. Firstly,
the placement of services directly
affects the resources they require
and the scheduling of workloads,
while resource provisioning
should match the requirements
of the services and their locations.
This indicates a critical need 
to balance service placement,
resource provisioning, and workloads scheduling.
Furthermore,
the popularity of services constantly
fluctuates over time and space,
and workload requests
may exhibit significant
short-term and long-term variations
, which undoubtedly further increase the difficulty of joint optimization.
The conventional approach is
to over-provision resources to ensure service performance,
but this 
can lead to high costs
and low resource utilization.
Therefore,
it is essential for service operators
to achieve efficient
resource allocation for services.
Finally, 
previous studies have frequently overlooked the essential requirement of managing system components across different timescales.
In a real system, workload requests (e.g., HTTP) are easily transferred to cloud servers or edge nodes for processing due to their small data size. However, services placement may not be adjusted fast enough to meet the
dynamic requirements of tasks.
For example, configuring services on a new server refers to downloading the application container (including libraries and data stores) from the cloud server, adaptive configurations, initializing the service, etc, which requires a certain amount of time and additional costs. Therefore, it is natural to manage services and workload requests across two distinct timescales, i.e., service placement and resource provisioning can be handled on a larger timescale, whereas workload request scheduling can take place on a smaller timescale. 
This method not only reduces placement costs but also decreases the frequency of predicting future workloads.
Multi-timescales approaches have demonstrated greater efficiency compared to single
timescale methods in edge computing environments~\cite{farhadi2021service}.

In this study,
we tackle the issue above
by devising a novel cloud-assisted
mobile edge computing framework
for dynamic resource management and workloads scheduling (RMWS).
Firstly,
we formulate the joint optimisation problem
of service placement,
resource provisioning
and workloads scheduling
to achieve minimised response time.
Theoretically,
the problem can be classified
as a Mixed Integer Non-Linear Programming (MINLP) problem,
which is difficult to solve directly.
Then considering the complexity
of the problem and the inherent differences
in the optimisation periods
of the three sub-problems~\cite{fan2022collaborative},
we have developed a two-timescale framework
that aims to effectively address these issues
by accounting for variations
in their respective optimization frequencies.
Besides, since service providers need to make a trade-off between cost and performance, our approach considers a scenario
under budget constraints
(including costs associated with
storage and computation resources)
that enable efficient allocation
of computation and storage resources.
Our contributions are summarized as follows:
%\begin{inparaenum}
\begin{itemize}
\item{We investigate service placement,
    resource provisioning and workloads scheduling
    in a cloud-assisted mobile edge computing system.
     Then use queuing theory to characterize the response time of workloads 
     and formulate the optimization goal as a mixed integer non-linear programming problem.}
    
\item{We develop a joint optimisation framework
    under two-timescales
    to solve the MINLP problem.
    On large timescales,
    we employ a Gibbs sampling algorithm 
    and an alternating minimisation method
    to determine the service placement and resource
    provisioning.
    On small timescales,
    we design a sub-gradient descent method
    to solve the workload scheduling challenge.}

\item{We conduct comprehensive 
    numerical simulations
    to evaluate the performance of the RMWS approach.
    The experiment results clearly demonstrate
    the superiority of our approach
    when compared with advanced methods.}
%\end{inparaenum}
\end{itemize}

% The remainder of the paper is structured as follows: Section II provides an overview of the related work. In Section III, we present the system model and formally define the problem. Next, in Section IV, we introduce our resource management and workload scheduling scheme. Section V performs simulations,
% followed by the conclusion in Section VI.
\section{Related work}

Recently, there has been a significant focus on enhancing the enforcement of service placement, resource allocation and computation offloading within edge-edge or edge-cloud collaboration environments.
Based on the optimization with different timescales, we have categorized the relevant work into two buckets: single timescale and multiple timescales.

\textbf{Joint Optimization on the Single Timescale.} Edge-cloud collaboration is
considered as an effective treatment
to solve the problem of
insufficient resources~\cite{
liu2022deep,ren2019collaborative, chu2022online}.
The authors in~\cite{liu2022deep}
presented a mechanism which utilizes
parameterized deep Q network
to jointly optimize service placement
and computational resource allocation
for task latency reduction.
% chu et al\cite{chu2022online} investigated the problem of maximizing user QoE in a resource-constrained MEC system through the joint optimization of service selection, resource allocation, and task offloading decisions. Considering the NP-hard nature of the problem, they reconfigured it into a Network Utility Maximization (NUM) problem and proposed a heuristic method based on resource efficiency.
In~\cite{ren2019collaborative},
the authors focused on the limited resources
in edge-cloud environments and proposed
an optimisation method
for joint communication and
computational resource allocation.

There are also some studies focusing on
Edge-Edge Collaboration~\cite{ chen2018task,huang2021price}.
% In \cite{ray2022prioritized}, Ray et al proposed an adaptive service placement policy that combines static and dynamic approaches. Utilizing Probabilistic Model Checking, they ensured probabilistic guarantees regarding the trade-offs between latency and energy consumption at edge sites. The static policy prioritized service placement, while the dynamic policy adjusted service placement configurations to accommodate user runtime variations.
In~\cite{chen2018task},
the authors explored the task offloading issue
in ultra-dense networks,
aiming to reduce delay
while preserving the battery life
of user equipment.
The work~\cite{huang2021price}
examined service placement
from the perspective of the service provider,
aiming to minimize the expenses
associated with service deployment
in the hierarchical mobile edge computing network
while fulfilling user requirements.

In reality,
taking into account both horizontal cooperation
among multiple edge nodes and vertical cooperation
between edge nodes and cloud servers
tends to result in more significant improvements
in system performance.
Therefore, researchers have actively
explored various potentials
in cloud-assisted
mobile edge computing systems~\cite{hao2020deep,poularakis2019joint}. 
The authors in~\cite{hao2020deep}
addressed the joint optimisation
problem under uncertain demand
in industrial cyber-physical system.
They proposed an enhanced DQN algorithm
for service placement and workload scheduling,
and utilized convex optimization techniques
to address resource allocation challenges.
Poularakis et al~\cite{poularakis2019joint}
designed a service placement
and request scheduling scheme with
multidimensional (storage-computation-communication)
constraints on the edge cloud.
% Ma et al.~\cite{ma2021dynamic}
% formulated the dynamic task scheduling problem
% in cloud-assisted mobile edge computing
% and proposed the WiDaS algorithm
% to deal with the challenges
% of task arrival dynamics,
% edge node heterogeneity
% and computation-communication delay trade-off.

However, in the above works, some issues are well-suited for optimization on a single timescale, such as task scheduling and task resource allocation. Nevertheless, regarding service deployment problems, we contend that it is not reasonable to approach service deployment and workloads scheduling optimization problems on the same timescale.
On one hand, there is a significant difference in the amount of data between workloads and services, which results in varying levels of difficulty in their transmission and distribution. On the other hand, there is also a discrepancy in the initialization and startup times between them.
If the deployment of services and resource allocation occur as frequently as the workloads scheduling, it can result in increased operational costs and heighten the instability of the system.

\textbf{Joint Optimization across Multi-timescales.}
Multi-timescales optimal design
has demonstrated greater efficacy
in fulfilling practical requirements,
with a scarcity of corresponding studies~\cite{farhadi2021service,li2021cooperative,wei2021joint}. 
The work~\cite{farhadi2021service} proposed a two-timescale framework for joint service placement and request scheduling under
resource constraints.
Considering user deadline preference and edge clouds’ strategic behaviors, Li et al.\cite{li2021cooperative} designed a novel perspective on cost reduction by exploiting the spatial-temporal diversities in workload
 and resource cost among federated Edge Clouds. Wei et al.\cite{wei2021joint} studied the joint optimization problem of resource placement and task dispatching in edge clouds across multiple timescales under the dynamic status of edge servers.

 In above studies, when considering the allocation of resources for service deployment, there is a common tendency to assume fixed CPU and storage configurations for the services. However, there is a lack of consideration from the perspective of the service provider in dynamically optimizing and adjusting the allocation of computation resources for services to meet the latency requirements of time sensitive tasks.
 In addition, a critical
factor that has been overlooked is the budget constraints for service providers to purchase server resources for service deployment.
Therefore,
we shift our focus to a full exploration
of joint optimisation problems across timescales in our research.

\section{System Model and Problem Formulation}
\subsection{System Overview}
We consider a typical cloud-assisted mobile edge computing system depicted in~\Cref{fig1}, consisting of IoT devices, edge servers, and a remote cloud server.
Define the set of edge servers as
$\mathcal{L} = \{1,2,...,i,...,L\}$
 and the cloud server is indexed as $L+1$.
Each edge server $i$
has a maximal computation capacity $F_{i}$
and storage capacity $M_{i}$.
The price of all storage and computation resources
in edge server $i$ are defined as $P_{i}^{m}$ and $P_{i}^{f}$.
The set of all possible services is
$\mathcal{S} = \{1,2,...,s,...,S\}$.
For each service $s$,
it needs an amount of storage resources
$m_{s}$ to cache related data which including software,
databases and models, etc.
We assume that $c_{s}$ is
the required CPU cycles
to process corresponding tasks.
Each service can be deployed
in a specific set of edge servers
and each edge server
can also host various services
depending on its available resources.
In addition, 
all services will be
deployed in the cloud.
We use a Poisson process to characterize the dynamics of task arrivals at edge nodes with the arriving rate $n_{i,s}$~\cite{ma2021dynamic}.
Then the total number of task requests
for service $s$
is noted as $n_{s} = \sum_{i=1}^{L}n_{i,s}$.

\textbf{{Service Placement Problem}}:
Let binary variable $x_{i,s}\in\{0,1\}$
indicator whether service $s$
is placed at edge server $i$.
The service placement decisions set
is denoted as
$\mathcal{X}=\{x_{i,s} | i\in\mathcal{L},s \in \mathcal{S}\}$.
Since the limited storage resource on edge node,
the aservice placements at edge server $i$
can not exceed the storage constrains:
 $\sum_{s=1}^{S}x_{i,s}m_{s} \leq M_{i}$.
% \begin{flalign}
% \sum_{s=1}^{S}x_{i,s}m_{s} \leq M_{i}, \quad \forall i\in\mathcal{L}.
% \end{flalign}

 \textbf{{Resource Provisioning Problem}}:
Let $y_{i,s}\in [0,1]$ represent
 the fraction of computation capacity
allocated to service $s$. The resource allocation set is denoted as
 $\mathcal{Y}=\{y_{i,s} | i\in\mathcal{L},
 s \in \mathcal{S}\}$. Here, $ \sum_{s=1}^{S}y_{i,s} \leq 1$.
Considering the budget constraints
$P_{i}^{bud}$ of edge server $i$  , $p_{i} = \sum_{s=1}^{S}x_{i,s}\bigg(\frac{m_{s}}{M_{i}}p_{i}^{m}+y_{i,s}P_{i}^{f}\bigg) \leq P_{i}^{bud}$.

%  \textbf{{Resource Provisioning Problem}}:
%  Let $y_{i,s}\in [0,1]$ represents
%  the fraction of computation resource
%  for service $s$ allocated by edge server $i$.
%  Accordingly, the resource allocation set
%  is denoted as
%  $\mathcal{Y}=\{y_{i,s} | i\in\mathcal{L},
%  s \in \mathcal{S}\}$.
%  Given the constrained computation resources
%  of edge servers,
%  the allocation of computation resources
%  must not surpass the constraint:
%  \vspace{-5pt}
% \begin{flalign}
% \sum_{s=1}^{S}y_{i,s} \leq 1, \quad \forall i\in\mathcal{L}.
% \end{flalign}
% Considering the storage and computation costs
% of services,
% as well as the budget constraints
% set by the service provider,
% it is essential to ensure that
% the total cost of resource allocation
% does not exceed the allocated budget $P_{i}^{bud}$:
% \begin{flalign}
% p_{i} = \sum_{s=1}^{S}x_{i,s}\bigg(\frac{m_{s}}{M_{i}}p_{i}^{m}+y_{i,s}P_{i}^{f}\bigg) \leq P_{i}^{bud}.
% \end{flalign}

\textbf{{Workload Scheduling Problem}}:
 Let $z_{i,s}\in[0,1]$ represent the workload ratio
 of the service $s$ 
 executed at edge service $i$.
 The workload scheduling policy
 is denoted as
 $\mathcal{Z}=\{z_{i,s} | i\in(\mathcal{L}\cup L+1),s \in \mathcal{S}\}$.
 For each service $s$,
 the workload ratio needs to
 satisfied the following constrains: 
  $\sum_{i=1}^{L+1}z_{i,s} = 1$.
%  \begin{flalign}
% \sum_{i=1}^{L+1}z_{i,s} = 1,  \quad \forall s \in \mathcal{S}.
% \end{flalign}
% \vspace{-10pt}
\subsection{Latency Model}
% Response latency is one
% of the most important metrics
% for evaluating the workload performance.
The latency described in our model
consists of both transmission latency
and computation latency. 

\textbf{{The Response Latency of Edge Servers:}}
Since IoT devices are usually close
to the wireless access point,
we ignore the transmission latency
between IoT devices and edge servers~\cite{8737385}.
Assuming the transmission latency
of the task served by service $s$
between edge servers
is $\phi_{s}$,
the transmission latency $T_{i,s}^{tran}$ can be obtained as:
\begin{flalign}
    T_{i,s}^{tran} = max\{z_{i,s}n_{s}-n_{i,s}, 0\}\cdot\phi_{s}.
\end{flalign}

% The computation latency
% is defined as the time from the task arriving
% to task completion
% and it consists of both waiting latency
% and execution latency.
We model the execution of tasks
on the edge server as an M/M/1 queue
and the computation latency $T_{i,s}^{com}$ of tasks
served by service $s$
which processed in edge server $i$ can be computed as:
\begin{flalign}
    T_{i,s}^{com} = \frac{z_{i,s}n_{s}}{y_{i,s}F_{i}/c_{s}-z_{i,s}n_{s}/\Delta t}.
\end{flalign}
To ensure the stability
of the queue or avoid infinite queue length,
according to queue theory we have:
\begin{flalign}
    \frac{y_{i,s}F_{i}}{c_{s}} - \frac{z_{i,s}n_{s}}{\Delta t} > 0, \quad \forall i\in\mathcal{L}, \quad \forall s \in \mathcal{S}.
\end{flalign}
In brief,
the total service response latency 
of the task offloading to the edge servers
can be expressed as:
\begin{flalign}
    T_{i} =  \sum_{s\in \theta_{i}}(T_{i,s}^{com}+T_{i,s}^{tran}),      
\end{flalign}
where $\theta_{i}$ denotes
the set of services placed on edge server $i$.

\textbf{{The Response Latency of Cloud Servers:}}
Given a well-supplied cloud data center with sufficient computation resources and all required services, our focus is solely on the transmission latency for offloading tasks to the cloud server, disregarding computation latency~\cite{liu2022deep,jie2021dqn}.
Let $\phi_{c,s}$ be the transmission delay
for the task served by service $s$
transferring to the cloud server.
Therefore the total transmission latency
of the task served by service $s$
offloading to the cloud server can be obtained as:
\vspace{-5pt}
\begin{flalign}
T_{c,s}^{tran} = z_{L+1,s}n_{s}\phi_{c,s}.
\end{flalign}

The total response latency
of the task offloading to the cloud server
can be obtained as:
\vspace{-5pt}
\begin{flalign}
T_{c} =\sum_{s=1}^{S}T_{c,s}^{tran}=\sum_{s=1}^{s}z_{L+1,s}n_{s}\phi_{c,s}.
\end{flalign}

\subsection{Problem Formulation}

In our work, we jointly optimized the edge service placement,
resource provisioning, and workloads scheduling policies,
aiming at minimizing response time.
Therefore, we formally define the problem as follows, denoted as \textbf{P1}. By observing, we identify it as a Mixed Integer Non-Linear Programming (MINLP) problem, which is challenging to solve directly.

\begin{flalign}
    \textbf{P1}:\quad  &\min_{\mathcal{X},\mathcal{Y},\mathcal{Z}}T_{total}= \min_{\mathcal{X},\mathcal{Y},\mathcal{Z}}\sum_{i=1}^{L} T_{i}+T_{c}   \\
     s.t. \quad &\sum_{s=1}^{S}x_{i,s}m_{s} \leq M_{i},  \quad \forall i\in\mathcal{L} \tag{C1}\\
     & \sum_{s=1}^{S}y_{i,s} \leq 1, \quad \forall i\in\mathcal{L}  \tag{C2}\\
       & \sum_{i=1}^{L+1}z_{i,s} = 1,\quad \forall s \in \mathcal{S} \tag{C3} \\
       & P_{i} \leq P_{i}^{bud}, \quad \forall i\in\mathcal{L}   \tag{C4} \\
          & z_{i,s}n_{s}c_{s}-y_{i,s}F_{i}\Delta t < 0, \quad \forall i\in\mathcal{L}, \quad \forall s \in \mathcal{S} \tag{C5} \\
          & x_{i,s} \in\{0,1\}, \quad \forall i\in\mathcal{L}, \quad \forall s \in \mathcal{S} \tag{C6} \\
          & y_{i,s} \in[0,1], \quad \forall i\in\mathcal{L}, \quad \forall s \in \mathcal{S} \tag{C7} \\
          & z_{i,s} \in [0,1], \quad \forall i\in\mathcal{L}, \quad \forall s \in \mathcal{S} \tag{C8}
\end{flalign}

% From the optimization problem P1,
% We observe that $\mathcal{X}$ is a binary integer variable,
% $\mathcal{Y}$ and $\mathcal{Z}$ are continuous variables.
% Thus, the problem P1 is a MINLP problem,
% which is NP-hard and difficult to solve directly.
\section{Resource Management and
Workloads Scheduling across Timescales}
\subsection{Problem Decoupling}

\begin{figure}[hb]
\centerline{\includegraphics[width=0.95\linewidth]{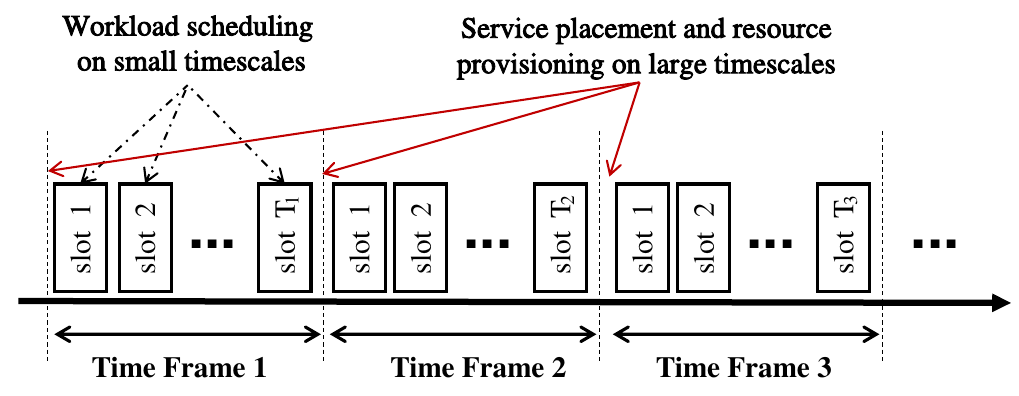}}
\caption{Joint optimization across two-timescale.}
\label{fig2}

\end{figure}

Considering the complexity of the problem
\textbf{P1},
we decoupled this problem and designed a two-timescale framework as displayed in~\Cref{fig2}.
The variables related to the time frame are denoted
with the superscript \textit{f} and subscript \textit{frame}
and those related to the time slot
are denoted with the superscript \textit{t} and subscript \textit{slot}.

At the start of time frame \textit{f},
we solve problem \textbf{P1} in an iterative manner with
the predicted request numbers
$n_{i,s} = n_{i,s}^{f}$, $n_{s} = n_{s}^{f}$
and $\Delta t = \Delta t^{f}$.
Note that only $\mathcal{X}_{frame}$ and $\mathcal{Y}_{frame}$ 
will be used in the time frame $f$, and
$\mathcal{Z}_{frame}$ will not be used for actual scheduling (named \textbf{shadow workload scheduling}),
but it plays an important role in assessing
the given service placement
and resource allocation strategies.
At the start of time slot \textit{t}, the request numbers
$n_{i,s} = n_{i,s}^{t}$, $n_{s} = n_{s}^{t}$
and $\Delta t = \Delta t^{t}$,
based on the optimal
$\mathcal{X}_{frame}^{*}$
and
$\mathcal{Y}_{frame}^{*}$
determined by time frame \textit{f},
we perform optimal workload scheduling
$\mathcal{Z}_{slot}^{*}$ in each time slot.

\begin{figure}[tbp]
\centerline{\includegraphics[width=0.9\linewidth]{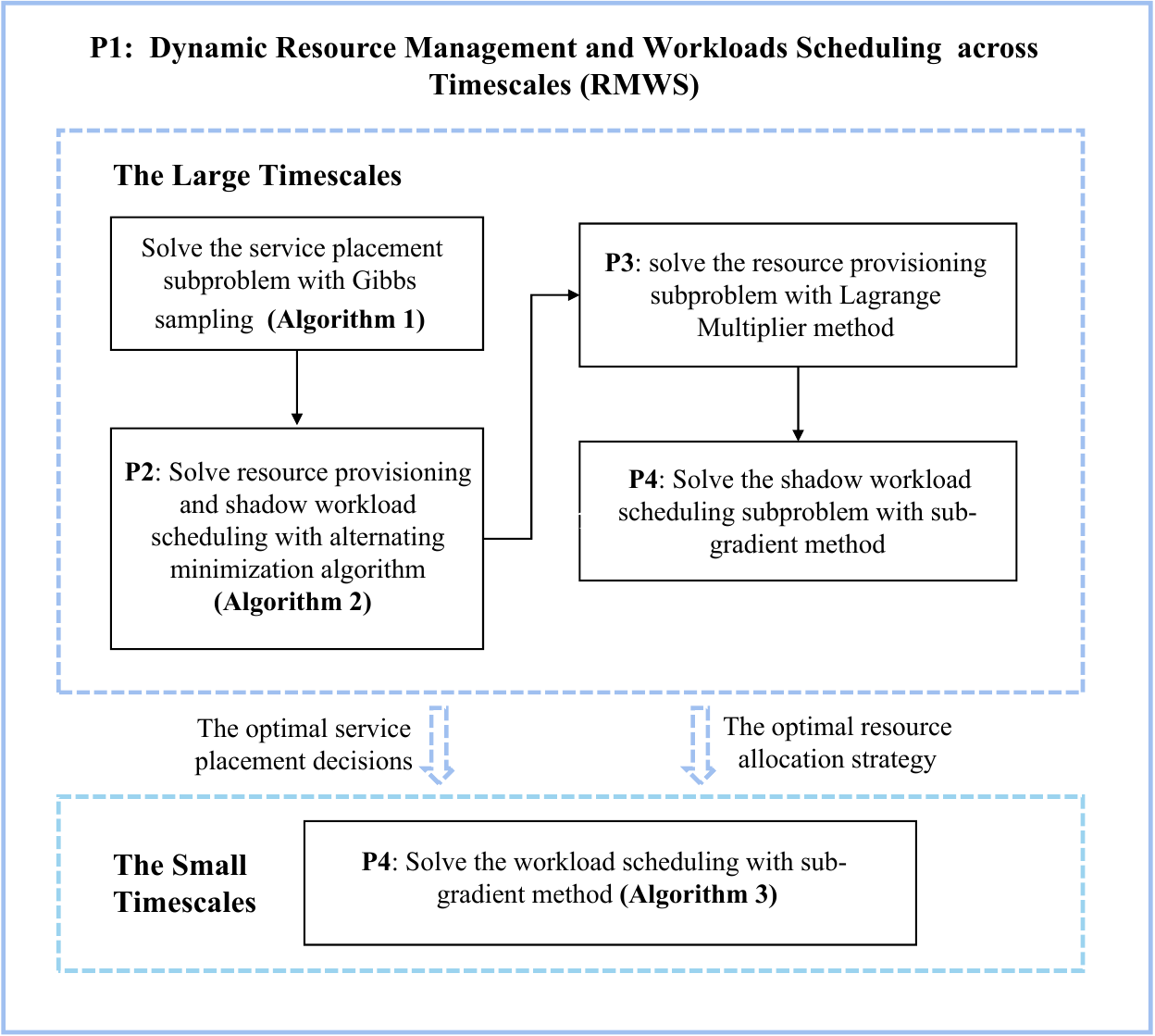}}
\caption{The flowchart of the RMWS algorithm}
\label{fig:RMWS}
\vspace{-10pt}
\end{figure}

On large timescales,
we decompose problem \textbf{P1} into several sub-problems and solve them using a two-layer iterative algorithm.
In the outer layer,
we update service placement decisions
based on Gibbs sampling \cite{zhou2021task}. 
In the inner layer,
 we solve resource provisioning
and shadow workload
scheduling with alternating
minimization algorithm.
On small timescales,
the service placement decisions
and resource allocation are given
and problem \textbf{P1} is reduced to
the workload scheduling problem.
we solve it based on
the sub-gradient method.
The flowchart of the RMWS
algorithm is depicted in \Cref{fig:RMWS}.

\subsection{service placement algorithm based on Gibbs sampling}
Gibbs sampling is a Markovian Monte Carlo method
commonly employed for dealing with
multidimensional random variables.
It can be introduced by the properties
of Markov chains
and transition probability matrix
to conclude that its sampling distribution
eventually converges to the joint distribution~\cite{gilks1995markov}.
The core concept of Gibbs sampling
is to randomly change one of
the variable while keeping the rest of
variables unchanged,
repeating the iterations until convergence.
% The algorithm can obtain
% a stationary distribution finally.

In the outer layer,
we use the idea of Gibbs sampling
to find the optimal service placement decisions
with objective value of \textbf{P1}.
We consider the process of
updating the service placement decisions
as a $L$-dimensional Markov chain
in which the $i$th dimension corresponds to
$i$th edge server service placement decision
and the algorithm is shown in \textbf{Algorithm 1}.
In each iteration,
the algorithm randomly selects an edge server \textit{k}
to update its service placement decision
$x_{k}$  to  $x_{k}^{\#}$
while ensuring that the rest edge servers
remain unchanged.
Thus the optimization problem \textbf{P1}
can be converted into
the following optimization problem \textbf{P2}
with the given service placement decisions
of all edge servers:
\begin{flalign}
    \textbf{P2}:\quad  &\min_{\mathcal{Y},\mathcal{Z}}T_{total}= \min_{\mathcal{Y},\mathcal{Z}}\sum_{i=1}^{L} T_{i}+T_{c},   \\
     s.t.& \quad C2-C5,\quad C7-C8 \notag 
\end{flalign}

The  problem \textbf{P2}
will be addressed in the \Cref{subsec:orp}.
And after solving it,
we can obtain the optimal objective value
$\vartheta$ 
(defined as 
$\vartheta = \min_{\mathcal{Y},\mathcal{Z}}T_{total}$).
The $\vartheta$ will change to $\vartheta^{\#}$
when the edge server $k$ updates 
decision $x_{k}$
to  $x_{k}^\#$ with the probability 
$\rho = \frac{1}{1+e^{(\vartheta^{\#}-\vartheta)/\omega}}$
and keep unchanged with probability $(1-\rho)$,
where $\omega$ is a smooth parameter and $\omega >0 $. 

\begin{algorithm}
\caption{Service Placement Based on Gibbs Sampling}\label{algorithm}
\renewcommand{\algorithmicrequire}{\textbf{Input:}}
\renewcommand{\algorithmicensure}{\textbf{Output:}}
\begin{algorithmic}[1]
    \REQUIRE   \quad\\$F_{i}$,  $M_{i}$, $p_{i}^{f}$, $p_{i}^{m}$, $p_{i}^{bud}$, $m_{s}$,  $c_{s}$, $n_{i,s}$  %%input
    \ENSURE \quad\\The optimal service placement decisions $\mathcal{X}_{frame}^{m}$ and resource allocation strategy $\mathcal{Y}_{frame}^{m}$ 
    \STATE Initialize service placement decisions $\mathcal{X}_{frame}^{0}$
    \FOR{each iteration $m$ = 1,2,...}
        \STATE Randomly select an edge server $k \in \mathcal{L}$ and an available service placement decision $x_{k}^{\#} \in \mathcal{X}_{k} $;
        \IF{$x_{k}^{\#}$ is feasible}
            
            \STATE Use \textbf{Algorithm 2} to compute $\mathcal{Y}_{frame}^{m-1}$, $\mathcal{Z}_{frame}^{m-1}$ and the corresponding $\vartheta$ by solving \textbf{P2}, based on $(x_{1}^{m-1},...,x_{k}^{m-1},...,x_{L}^{m-1})$.
            
            \STATE Use \textbf{Algorithm 2} to compute $\mathcal{Y}_{frame}^{\#}$, $\mathcal{Z}_{frame}^{\#}$ and the corresponding $\vartheta^{\#}$ by solving \textbf{P2}, based on $(x_{1}^{m-1},...,x_{k}^{\#},...,x_{L}^{m-1})$.

            \STATE Let $\rho = \frac{1}{1+e^{(\vartheta^{\#}-\vartheta)/\omega}}$.
            \STATE $x_{k}^{m}=x_{k}^{\#}$, $\mathcal{Y}_{frame}^{m} = \mathcal{Y}_{frame}^{\#}$, $\mathcal{Z}_{frame}^{m}=\mathcal{Z}_{frame}^{\#}$ with the probability $\rho$.
            
            \STATE $x_{k}^{m}=x_{k}^{m-1}$, $\mathcal{Y}_{frame}^{m} = \mathcal{Y}_{frame}^{m-1}$, $\mathcal{Z}_{frame}^{m}=\mathcal{Z}_{frame}^{m-1}$ with the probability $1-\rho$.
        \ENDIF
       \IF{the stopping criterion is satisfied}
            \STATE Return $\mathcal{X}_{frame}^{m}$, $\mathcal{Y}_{frame}^{m}$
       \ENDIF
    \ENDFOR
\end{algorithmic}
\end{algorithm}

\textbf{Theorem 1.} 
As the value of  $\omega$ decreases,
\textit{Algorithm 1} is more likely
to converge to the global optimal for problem \textbf{P1}
and  will converge to the global optimal solution
with probability 1 when $\omega \rightarrow 0$.

\textbf{Proof.} Please see Appendix A.

\subsection{Optimal resource provisioning and shadow workload scheduling with alternating minimization algorithm}
\label{subsec:orp}

Once the service placement decisions
are determined in the outer layer,
we face the task of
resolving the resource allocation
and workload scheduling problem,
referred to as \textbf{P2} in the inner layer.
It can be proved that the optimization problem \textbf{P2}
is a non-convex problem with two variables $(y_{i,s},z_{i,s})$
and solving it directly remains a challenging task.
Thus, we propose an alternating minimization algorithm,
whose detailed procedure is shown in \textbf{Algorithm 2}. 

When we fix the variable $z_{i,s}$
as a constant value $z_{i,s}^{*}$,
then the problem \textbf{P2}
can be converted into the following problem \textbf{P3}
with respect to $y_{i,s}$.
it is easy to prove that this problem
is a convex optimization problem,
and classical Karush-Kuhn-Tucker (KKT) conditions~\cite{gordon2012karush}
can be applied to find a closed-form solution.
See \textit{Sub-section D} for more details.
\begin{flalign}
    \textbf{P3}:\quad  &\min_{\mathcal{Y}}T_{total}= \min_{\mathcal{Y}}\sum_{i=1}^{L}T_{i}+T_{c},   \\
     s.t.& \quad C2,\quad C4-C5,\quad C7 \notag 
\end{flalign}
Similarly, when we fix the variable $y_{i,s}$
as a constant value $y_{i,s}^{*}$,
then the optimization problem \textbf{P2}
can be converted into the following problem
\textbf{P4} with respect to $z_{i,s}$.
It is easy to prove that this problem
is a convex optimization problem,
and the sub-gradient descent method
can be applied to find the optimal solution.
See \textit{Sub-section E} for more details.

\begin{flalign}
    \textbf{P4}:\quad  &\min_{\mathcal{Z}}T_{total}= \min_{\mathcal{Z}}\sum_{i=1}^{L}T_{i}+T_{c}, \label{con:p4} \\
     s.t.& \quad C3,\quad C5,\quad C8 \notag 
\end{flalign}

\begin{algorithm}

\caption{Alternating Optimization-Based Algorithm}\label{algorithm2}
\renewcommand{\algorithmicrequire}{\textbf{Input:}}
\renewcommand{\algorithmicensure}{\textbf{Output:}}
\begin{algorithmic}[1]
    \REQUIRE   \quad\\The service placement decisions $\mathcal{X}_{frame}$, task request numbers in time frame $n_{i,s}^{f}$, error tolerance threshold $\epsilon$  %%input
    
    \ENSURE \quad\\The optimal resource allocation strategy $\mathcal{Y}_{frame}^{n}$ and shadow workload scheduling $\mathcal{Z}_{frame}^{n}$ under the service placement decisions $\mathcal{X}_{frame}$. %%output
   
    \STATE Initialize $\mathcal{Z}_{frame}^{0}$.
    \STATE Obtain  $\mathcal{Y}_{frame}^{0}$ based on the Lagrange multiplier method.
    \STATE Calculate $f(\mathcal{Z}_{frame}^{0})$ and $g^{(0)}$ according to Eq.(\ref{con:p4}) and Eq.(\ref{con:9})
    \FOR{each iteration $n$ = 1,2,...}
        \STATE Update the shadow workload scheduling vector by Eq. (\ref{con:8}) and perform the weighting operation to satisfy constraint C3 and C8.
        \STATE Obtain resource allocation strategy $\mathcal{Y}_{frame}^{n}$ based on the Lagrange multiplier method.
       \STATE Calculate $f(\mathcal{Z}_{frame}^{n})$ and $g^{(n)}$ according to Eq.(\ref{con:p4}) and Eq.(\ref{con:9})
       \IF{$|f(\mathcal{Z}_{frame}^{n}) - f(\mathcal{Z}_{frame}^{n-1})| \leq \epsilon$}
            \STATE Return $\mathcal{Y}_{frame}^{n}$, $\mathcal{Z}_{frame}^{n}$
       \ENDIF
    \ENDFOR
\end{algorithmic}
\end{algorithm}
\textbf{Theorem 2.}  Problem \textbf{P2} is a non-convex optimization problem for a given optimal service placement decisions $\mathcal{X}^{*}$.

\textbf{Proof.} Please see Appendix B.

\textbf{Theorem 3.} The \textbf{P3} and \textbf{P4} are the convex optimization problems.

\textbf{Proof.} Please see Appendix C.

\subsection{Optimal resource provisioning with Lagrange Multiplier method}
According to \textbf{Theorem 3}, 
problem \textbf{P3} is can be solved by the Lagrange multiplier method~\cite{rockafellar1993lagrange}.
The Lagrangian function for problem
\textbf{P3} is established as follows:

\begin{flalign}
L(f(\mathcal{Y}),\lambda,\mu) &=f(\mathcal{X^{*}},\mathcal{Z^{*}},\mathcal{Y})+\sum_{i=1}^{L}\lambda_{i}(\sum_{s\in \theta_{i}}y_{i,s}-1)\\&+\sum_{i=1}^{L}\mu_{i}\bigg[\sum_{s\in \theta_{i}}\bigg(\frac{m_{s}}{M_{i}}P_{i}^{m}+y_{i,s}P_{i}^{f}\bigg)-P_{i}^{bud}\bigg] \notag.
\end{flalign}
 where $\lambda$ and $\mu$
 are the Lagrangian multiplier.
 Then as shown in \textbf{Theorem 4},
 we can derive the optimal solution by using KKT conditions.

\textbf{Theorem 4.} The optimal solution of problem \textbf{P3} is given as follows.

\textit{Case 1}: if $\Gamma_{i} < 1$, the optimal solution is given by
\begin{flalign}
  & y_{i,s}^{*}=\frac{\sqrt{z_{i,s}^*n_{s}c_{s}}(\Gamma_{i}F_{i}\Delta t-\sum_{s\in\theta_{i}}z_{i,s}^*n_{s}c_{s})}{\sum_{s\in\theta_{i}}\sqrt{z_{i,s}^*n_{s}c_{s}}F_{i}\Delta t}+\frac{z_{i,s}^*n_{s}c_{s}}{F_{i}\Delta t}.
\end{flalign}

\textit{Case 2}: if $\Gamma_{i} \geq 1$, the optimal solution is given by
\begin{flalign}
    &y_{i,s}^{*}=\frac{\sqrt{z_{i,s}^*n_{s}c_{s}}(F_{i}\Delta t-\sum_{s\in\theta_{i}}z_{i,s}^*n_{s}c_{s})}{\sum_{s\in\theta_{i}}\sqrt{z_{i,s}^*n_{s}c_{s}}F_{i}\Delta t}+\frac{z_{i,s}^*n_{s}c_{s}}{F_{i}\Delta t},
\end{flalign}
where $\Gamma_{i}=\bigg(p_{i}^{bud}-\sum_{s\in \theta_{i}}\frac{m_{s}}{M_{i}}P_{i}^{m}\bigg)/P_{i}^{f}$.

\textbf{Proof.} Please see Appendix D.

\subsection{Optimal shadow workload scheduling with sub-gradient method}
The key challenge of Problem \textbf{P4}
is that Eq. (\ref{con:p4}) is continuous but non-differential
(or non-smooth) at $z_{i,s}n_{s} = n_{i,s}$.
As a result, it is difficult to use the traditional
KKT conditions directly and find a closed-form solution.
For non-differential convex problems,
a common approach is the
sub-gradient descent method~\cite{2004Convex}.
Therefore,
we design an efficient algorithm
based on sub-gradient descent.
The sub-gradient of optimization objective \textbf{P4} is represented as
: $\frac{\partial f(\mathcal{Z})}{\partial z_{i,s}}$

\begin{flalign}
    =
    \begin{cases}
   \frac{n_{s}c_{s}y_{i,s}^*F_{i}\Delta t^2}{(y_{i,s}^*F_{i}\Delta t-z_{i,s}n_{s}c_{s})^{2}}+n_{s}\phi_{s}  & z_{i,s}n_{s} > n_{i,s}, i \in \mathcal{L}  \\
    \bigg[ \frac{n_{s}c_{s}y_{i,s}^*F_{i}\Delta t^2}{(y_{i,s}^*F_{i}\Delta t-z_{i,s}n_{s}c_{s})^{2}},\\ \quad 
    n_{s}\phi_{s}+\frac{n_{s}c_{s}y_{i,s}^*F_{i}\Delta t^2}{(y_{i,s}^*F_{i}\Delta t-z_{i,s}n_{s}c_{s})^{2}} \bigg] &  z_{i,s}n_{s} = n_{i,s}, i \in \mathcal{L}  \\
    \frac{n_{s}c_{s}y_{i,s}^*F_{i}\Delta t^2}{(y_{i,s}^*F_{i}\Delta t-z_{i,s}n_{s}c_{s})^{2}} & z_{i,s}n_{s} < n_{i,s}, i \in \mathcal{L}  \\
    n_{s}\phi_{c,s} & i = L+1
    \end{cases} 
\end{flalign}

The problem \textbf{P4} can be solved by sub-gradient descent method in the following iteration:
\begin{flalign}
    z_{i,s}^{(n+1)} = z_{i,s}^{(n)}-\alpha_{n}g^{(n)}, \label{con:8}
\end{flalign}
where $z_{i,s}^{(n)}$
denotes the value of the $n$th iteration
of $z_{i,s}$,
$\alpha_{n} >0$ denotes the step size
of the $n$th iteration
and $g$ is the sub-gradient
of the function $f(\mathcal{Z})$
at $z_{i,s}^{(n)}$,
it can be written as
\begin{flalign}
g = \begin{cases}
    \partial f(\mathcal{Z}), & \text{subject to C5} \\
    \partial(z_{i,s}n_{s}c_{s}-y_{i,s}^{*}F_{i}\Delta t), & \text{if } (z_{i,s}n_{s}c_{s}-y_{i,s}^{*}F_{i}\Delta t) > 0.
\end{cases} \label{con:9}
\end{flalign}
Notice that the constraints C3 and C8
will be satisfied by weighted operation,
and $\partial(z_{i,s}n_{s}c_{s}-y_{i,s}^{*}F_{i}\Delta t)$
is utilized as the obstacle function.

% According to the above
% \textbf{Theorem 5}, the problem \textbf{P4}
% can be efficiently solved by iteratively
% updating the shadow workload scheduling.
% The details are shown in \textbf{Algorithm 2}.

\subsection{Workloads Scheduling on Small Timescales}
So far,  we have discussed how to deploy service placement and allocate resources on a large timescale. The next matter that needs to be determined is the workload scheduling problem for each time slot.
The optimal service placement decisions $\mathcal{X}_{frame}^*$ and
 resource allocation $\mathcal{Y}_{frame}^*$ will be given at the start of every time frame \textit{f}. Then the problem \textbf{P1} will be simplified to the problem \textbf{P4} at the start of every time slot \textit{t}. It can also be solved by sub-gradient method, which is summarized in \textbf{Algorithm 3}.

\begin{algorithm}

\caption{Workloads scheduling on small timescales}\label{algorithm3}

\renewcommand{\algorithmicrequire}{\textbf{Input:}}
\renewcommand{\algorithmicensure}{\textbf{Output:}}
\begin{algorithmic}[1]
    \REQUIRE   \quad\\ $\mathcal{X}_{frame}^{*}$ and $\mathcal{Y}_{frame}^{*}$ in the time frame $f$, service request numbers $n_{i,s}^{t}$ in the time slot $t$.
    
    \ENSURE \quad\\The optimal workload scheduling $\mathcal{Z}_{slot}^{n}$ in the time slot $t$. %%output
   \STATE Set service placement $\mathcal{X}_{frame}^{*}$ and resource allocation strategy $\mathcal{Y}_{frame}^{*}$. 
    \STATE Initialize the workload scheduling $\mathcal{Z}_{slot}^{0}$.
    \STATE Calculate $f(\mathcal{Z}_{slot}^{0})$ and $g^{(0)}$ according to Eq.(\ref{con:p4}) and Eq.(\ref{con:9})
    \FOR{each iteration $n$ = 1,2,...}
        \STATE Update the workload scheduling vector by Eq. (\ref{con:8}) 
       \STATE Calculate $f(\mathcal{Z}_{slot}^{n})$ and $g^{(n)}$ according to Eq.(\ref{con:p4}) and Eq.(\ref{con:9})
       \IF{$|f(\mathcal{Z}_{slot}^{n}) - f(\mathcal{Z}_{slot}^{n-1})| \leq \epsilon$}
            \STATE Return $\mathcal{Z}_{slot}^{n}$
       \ENDIF
    \ENDFOR
      
\end{algorithmic}
\end{algorithm}

\textbf{The complexity
of RMWS}:
In \textbf{P3}, 
$y_{i,s}^*$ can be computed by
the closed-form expressions,
whose complexity can be neglected.
In \textbf{P4},
the complexity of the sub-gradient descent algorithm
mainly related to the number of iterations.
Based on the proof in \cite{2004Convex},
it has a polynomial time complexity of $O(1/\epsilon^2)$.
Therefore, the overall complexity
of \textbf{Algorithm 3} and \textbf{Algorithm 2} is $O(1/\epsilon^2)$.
The complexity of \textbf{Algorithm 1} is highly correlated with the number of Gibbs sampling iterations $M$ and the complexity of \textbf{Algorithm 2}, which is 
$O(M/\epsilon^2)$. 
\section{Performance Evaluations}

\begin{figure*}[h]
  \centering
  % 第一张图片
  \begin{minipage}[b]{0.31\textwidth}
    \includegraphics[width=\textwidth]{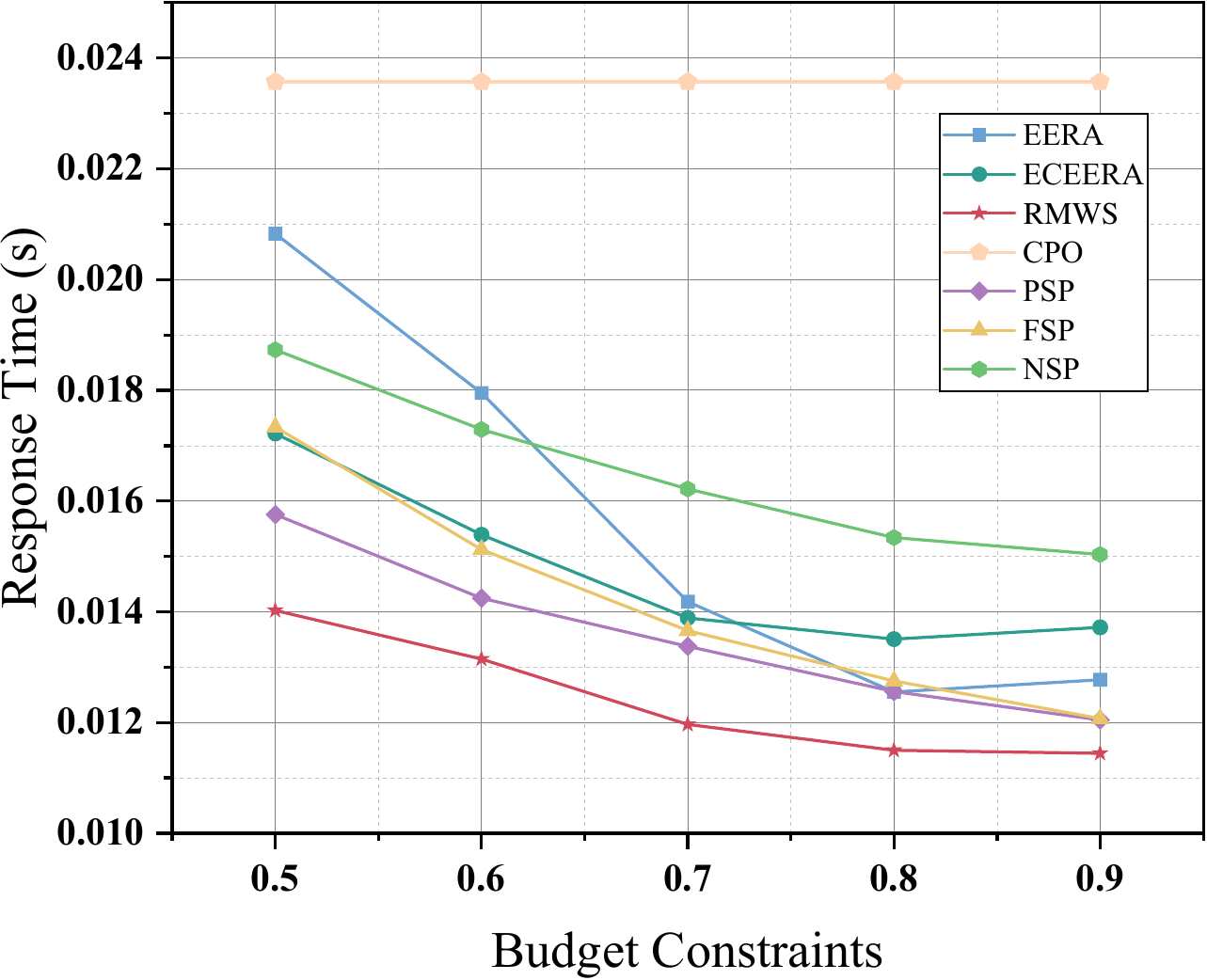}
    \caption{Response time under different budget constraints}
    \label{budget constraints}
  \end{minipage}
  \quad 
  % 第二张图片
  \begin{minipage}[b]{0.31\textwidth}
    \includegraphics[width=\textwidth]{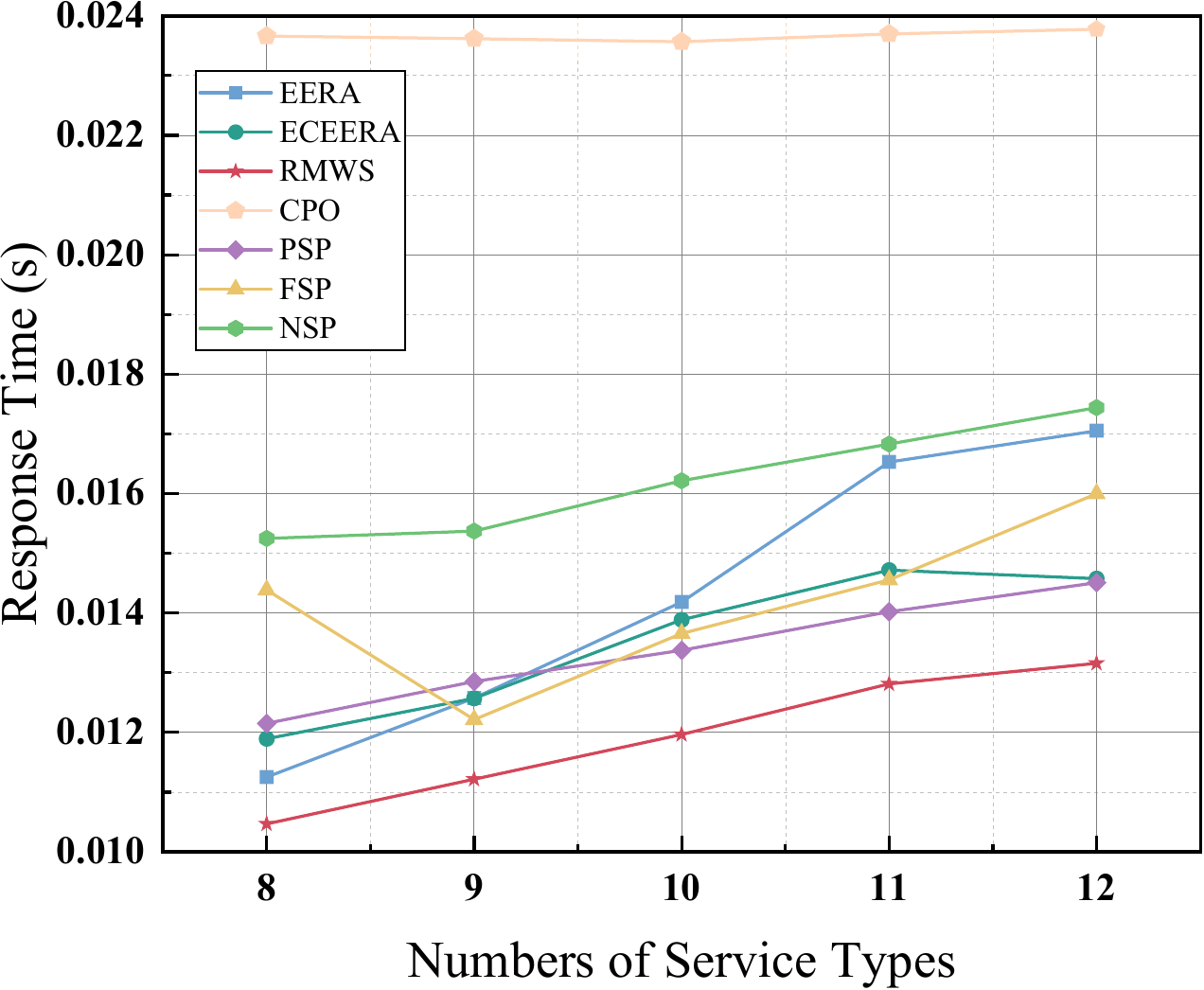}
    \caption{Response time under different numbers of service types}
     \label{Numbers_of_Service_Types}
  \end{minipage}
 \quad
  \begin{minipage}[b]{0.31\textwidth}
    \includegraphics[width=\textwidth]{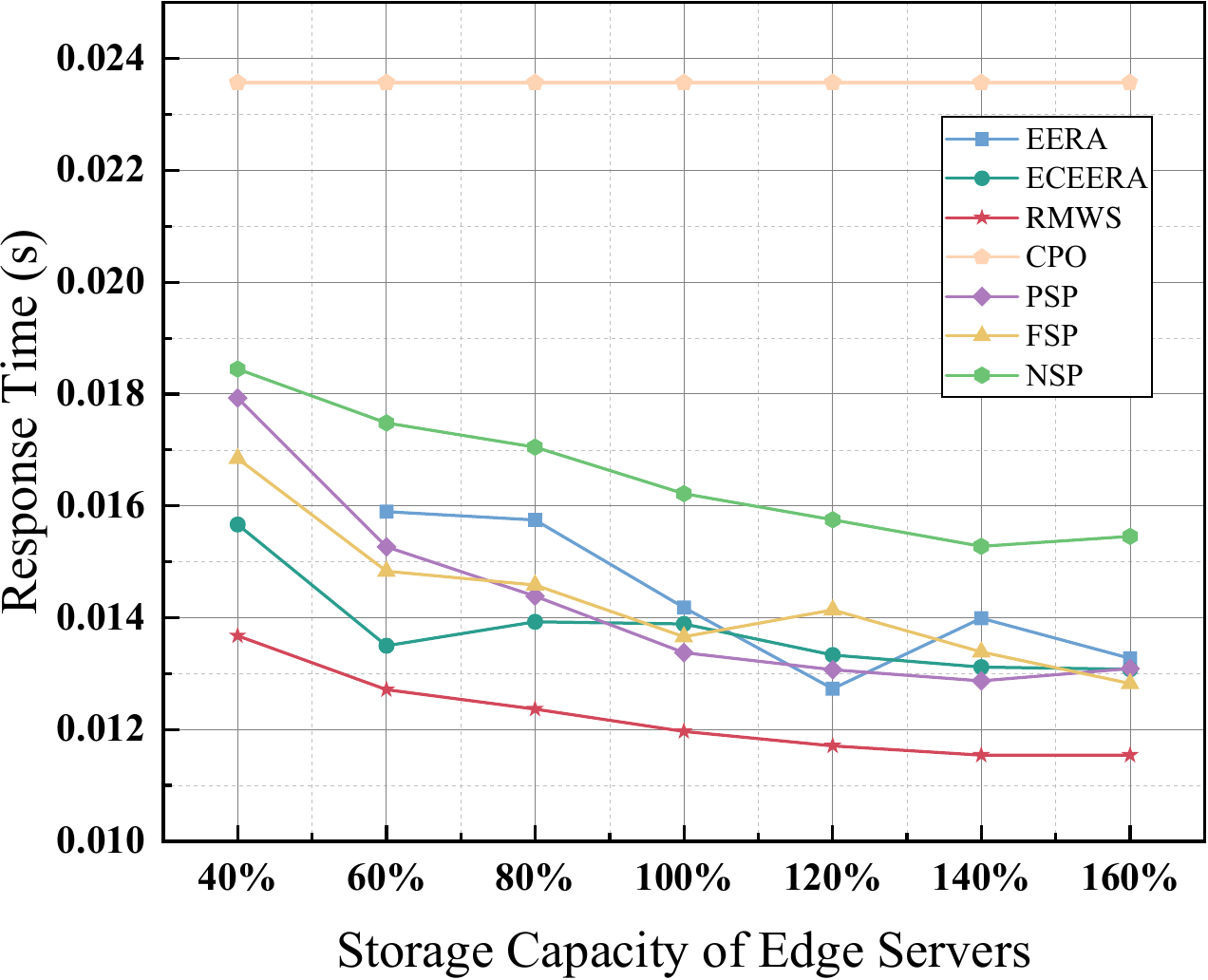}
    \caption{Response time under different storage capacity of edge servers.}
        \label{edge_storage_capacity}
  \end{minipage}
 
\end{figure*}

\subsection{Experiment Setup}
{\textbf{{Parameter Setting:}}
The default settings of parameters
in our simulations are collected
in~\Cref{tab:sim}.
We note that the values chosen
for the parameters are practical
and are widely used in existing work~\cite{huang2021price,
ma2020cooperative,li2022joint}
and the price of edge server resource
refers to Alibaba Cloud servers\footnote{https://www.aliyun.com/price}.}

\begin{table}[htp] %%参数： h:放在此处 t:放在顶端 b:放在底端 p:在本页
	\renewcommand\arraystretch{1.2}

	\centering  % 显示位置为中间

	\caption{SIMULATION SETUP AND SYSTEM PARAMETERS\label{tab:sim}}
    \setlength{\tabcolsep}{1mm}{

	\begin{tabular}{p{160pt}l|p{70pt}l} %第一列设置宽度为45pt 全为左对齐 没有分割线
		\hline  % 表格的横线
		\toprule % 顶部线
		\textbf{Parameters} & & \textbf{Values}  \\
		\hline  % 表格的横线
		\hline
		%\midrule % 中部线
		Number of edge servers, $L$  & & 4          \\  
		Number of services, $S$  & & 10          \\ 
		Edge server storage resource, $M_{i}$ & & [50,200]GB  \\
		Edge server computation resource, $F_{i}$ & & [50,150] GHz \\
	    Service storage requirement, $m_{s}$ & & [10,40]GB \\
	    Service computation requirement,  $c_{s}$ & & [0.1,0.5] GHz \\
	    Edge server storage resource price, $P_{i}^{m}$ & & [10, 40] CNY/hour \\
	    Edge server computation resource price, $P_{i}^{f}$ & & [10,50] CNY/hour \\
	    zipf distribution coefficient, $e$ & & 0.6 \\
	    budget coefficient, $\mu$ & & 0.7  \\
	    Smooth parameter, $\omega$ & & 0.001 \\
		%\midrule

		\bottomrule % 底部线
		%\hline  % 表格的横线
	\end{tabular}}
\end{table}

{\textbf{{Service Request Demand:}}
We assume that the service request demands
of each edge server follow Zipf distribution,
witch are consistent with
the other researches~\cite{li2022joint}.
The Zipf's law shows that the probability
for a single request to
the type \textit{s} service is
$p_{s} = \frac{1}{s^e H}$ for all services,
i.e. $p_{s}$ is the popularity of service \textit{s},
where $H = \sum_{s=1}^{|S|}\frac{1}{s^{e}}$, $1 \leq s \leq |S|$
and these services are ranked in their popularity.}

{\textbf{{Task Arrival Pattern:}}
Mobile edge computing has not been
widely deployed in practice,
thus we can not trace the mobile task arrivals
at edge server precisely.
In our work, we use the mathematical traffic pattern
to simulate the time-varying property
of mobile task arrivals at edge servers~\cite{ma2021dynamic}.
Assume that each time frame contains
30 time slots and per slot length is one minute. In each time frame,
the number of task requests at each edge server obeys a normal distribution with a mean of 600 tasks/slot and a variance of 20 tasks/slot.

\subsection{ Benchmark Algorithms}
We evaluate the performance of our algorithm RMWS with following state-of-the-art algorithms:

% The algorithm proposed
% in this paper will be compared with some benchmark algorithms.
% In order to assess the performance of
% the proposed service placement algorithm,
% we use the following four benchmarks algorithms.
\begin{enumerate}
\item{{\textbf{Cloud Processing Only (CPO)}}: 
No services are placed on the edge servers
and all requested tasks are offloaded
to the cloud for processing.}

\item{{\textbf{Fixed Service Placement (FSP) Algorithm}}~\cite{naouri2021novel}:
Services placed on edge servers
are fixed without service placement optimization.}

\item{{\textbf{Non-cooperation Service Placement (NSP) Algorithm}}~\cite{liu2022deep}}:
Each edge server performs
independent service placement,
with the workload either processed
on local edge server or offloading
to the cloud server and
the edge-edge cooperation is disabled.

\item{{\textbf{Popular Service Placement (PSP) Algorithm}}\cite{li2023optimal}:
Services are placed on
edge servers according to popularity.
Workload scheduling and computation resource provisioning
are jointly optimized.}

\item{{\textbf{Edge-Edge Cooperation Scheduling
with Equal Resource Allocation (EERA)}}~\cite{ma2020leveraging}}:
which disables edge-cloud cooperation
and resource allocation optimization.

\item{{\textbf{ Edge-Cloud and Edge-Edge Cooperation Scheduling
with Equal Resource Allocation (ECEERA)}}~\cite{ma2020cooperative}}:
which disables resource allocation optimization.

\end{enumerate}

% To assess the performance
% of the proposed resources provisioning
% and workload scheduling algorithm,
% we use the following benchmarks:
% \begin{enumerate}
% \item{{\textbf{Edge-Edge Cooperation Scheduling
% with Equal Resource Allocation (EERA)}},
% which disables edge-cloud cooperation
% and resource allocation optimization.
% And optimal service placement and workload scheduling
% are conducted.
% The idea of this scheme
% is embodied in~\cite{ma2020leveraging}}
% \item{{\textbf{ Edge-Cloud and Edge-Edge Cooperation Scheduling
% with Equal Resource Allocation (ECEERA)}},
% which disables resource allocation optimization.
% And optimal service placement
% and workload scheduling are conducted.
% The idea of this scheme
% is embodied in~\cite{ma2020cooperative}}
% \end{enumerate}

\begin{figure*}[ht]
  \centering
  % 第一张图片
  \begin{minipage}[b]{0.31\textwidth}
  \includegraphics[width=\textwidth]{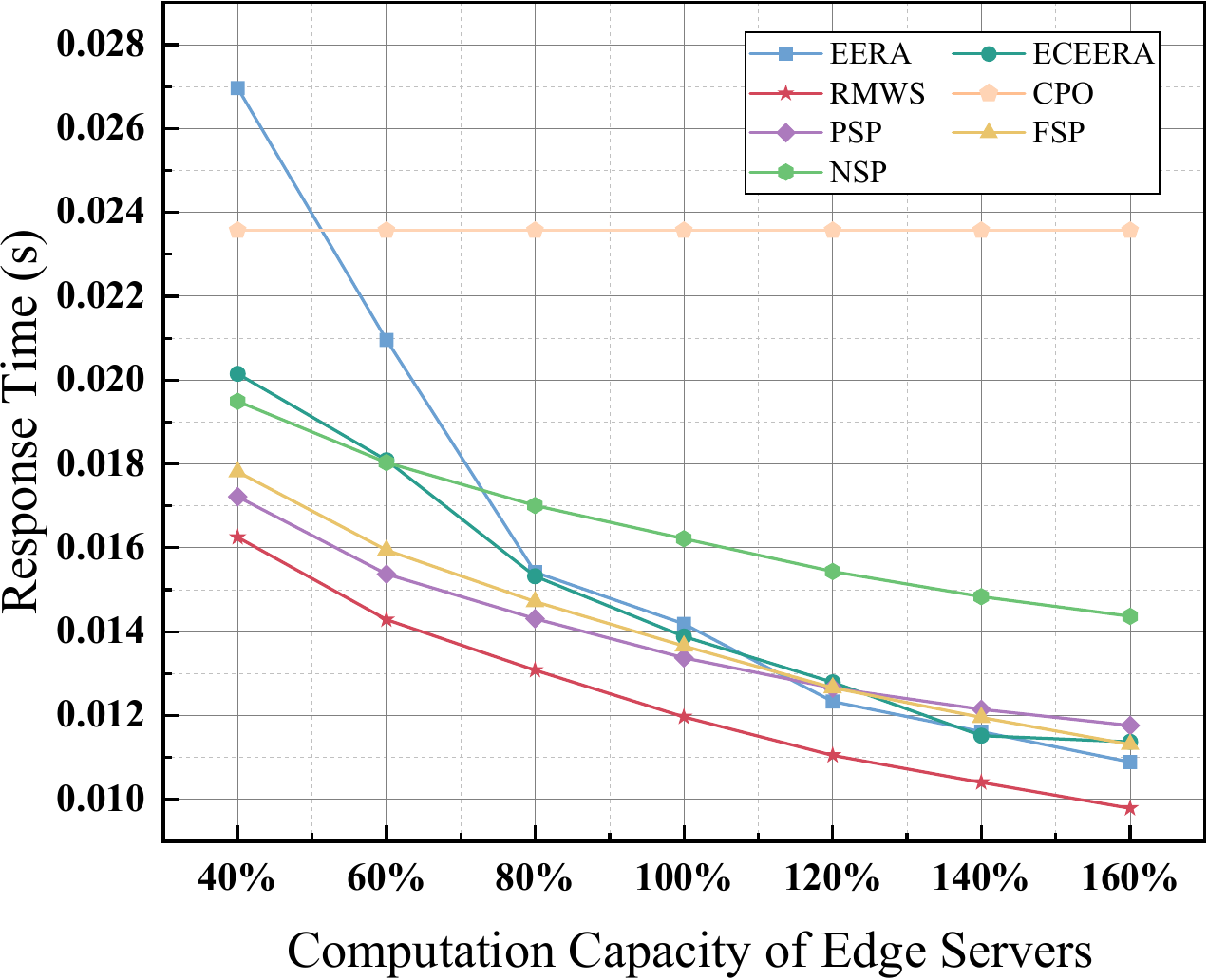}
   \caption{Response time under different computation capacity of edge servers.}
    \label{edge_computation_capacity}
  \end{minipage}
  \quad 
  % 第二张图片
  \begin{minipage}[b]{0.31\textwidth}
    \includegraphics[width=\textwidth]{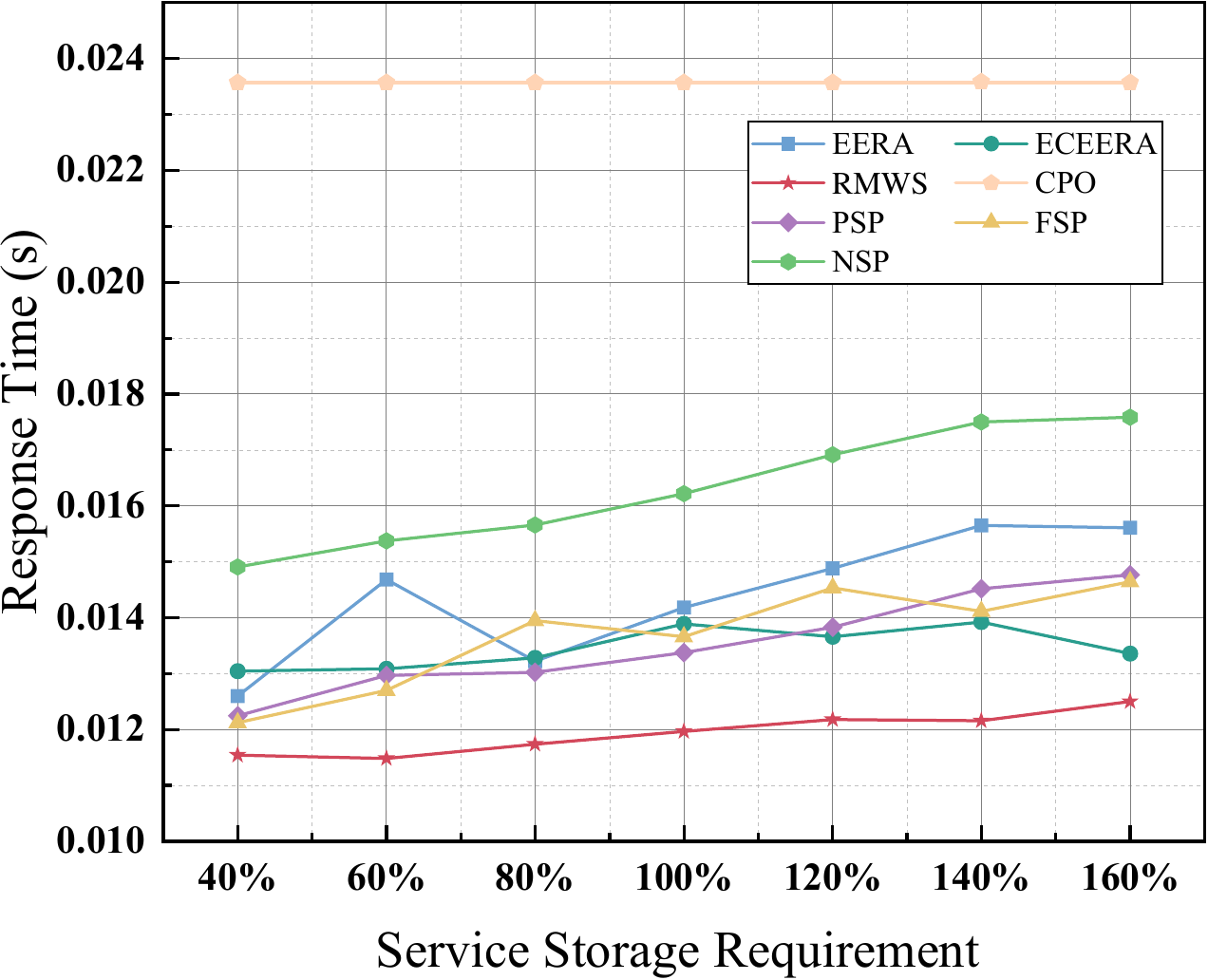}
    \caption{Response time under different storage requirement of services.}\label{service_storage_requirement}
  \end{minipage}
 \quad
  \begin{minipage}[b]{0.31\textwidth}
   \includegraphics[width=\textwidth]{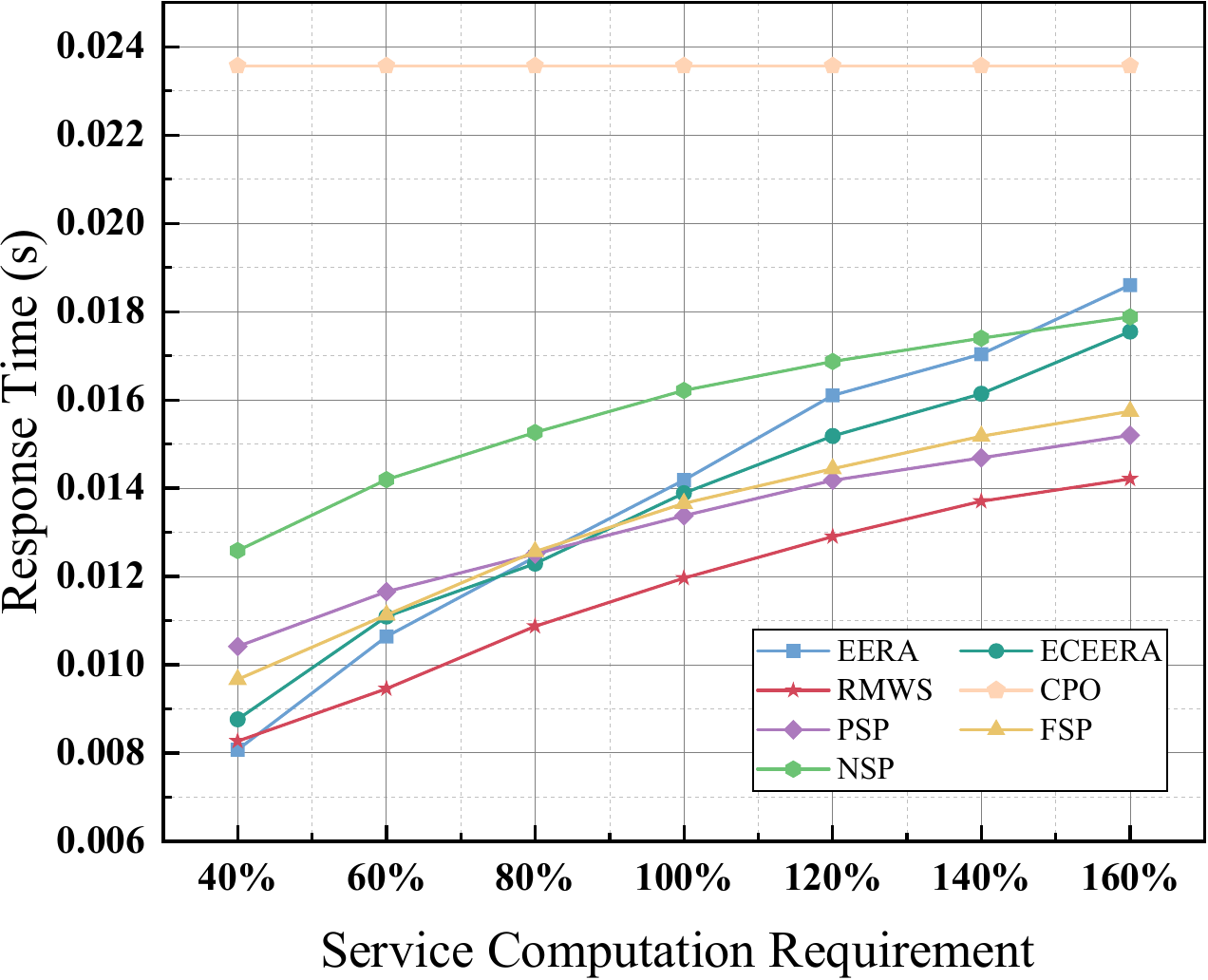}
    \caption{Response time under different computation requirement of services.}\label{service_computation_requirement}
  \end{minipage}
 
\end{figure*}

% \begin{figure}[ht]
% \centerline{\includegraphics[scale=0.3]{pic/budget_constraints.pdf}}
% \caption{Response time under different budget constraints}
% \label{budget constraints}
% \end{figure}
% \begin{figure}[ht]
% \centerline{\includegraphics[scale=0.3]{pic/Numbers_of_Service_Types.pdf}}
% \caption{Response time under different numbers of service types}
% \label{Numbers_of_Service_Types}
% \end{figure}

\subsection{Simulation Results}

\begin{inparaenum}
\item \textbf{Comparison under Different Budget Constraints.}
To evaluate the impact
of allocated budge,
% The workload response time
% is influenced by the budget allocated.
we enhance the budget constraint for each edge server,
raising it from 50\% to 90\%
of the total cost of computation
and storage resources across
all edge servers.
Meanwhile,
the remaining parameter values are kept constant.
It can be seen from ~\Cref{budget constraints},
the response latency of workloads
shows a tendency to decrease
with the elevation of the service provider's budget.
The expanded budget enhances
the availability of resources on edge servers,
facilitating the deployment of services
and processing of tasks.
Notably, 
the response time of the service
remains unaffected by the change in budget
when employing the CPO approach,
as the workloads are processed in the cloud.

    \item \textbf{Comparison under
Different Numbers of Service Types.}
% To evaluate the impact of service types,
% The placement of service will be influenced
% by the increase of service types.
% in~\Cref{Numbers_of_Service_Types},
% we increase the number of service types
% from 8 to 12 while keeping the other parameter values constant.
From the ~\Cref{Numbers_of_Service_Types},
we can see that as the number
of service types increases,
there is a gradual rise 
in the average response time for each algorithm.
This is primarily due to the limited resources
of the edge servers,
which cannot accommodate
the growing number of services for placement.
Tasks associated with services
that cannot be placed on the edge servers
are consequently offloaded to the cloud,
thus prolonging the task processing delay
caused by edge-cloud transmission.
When the number of applications is 12,
the RMWS reduces the response time
by 10.7\%-80.9\% compared to other algorithms. The main reason is that RWMS can more effectively allocate resources on edge servers to various services and coordinate response latency between edge servers and cloud servers.

\begin{figure*}[ht]
    \centering
    \subfigure[The variations of Zipf distribution coefficients.]{
        \includegraphics[width=0.45\textwidth]{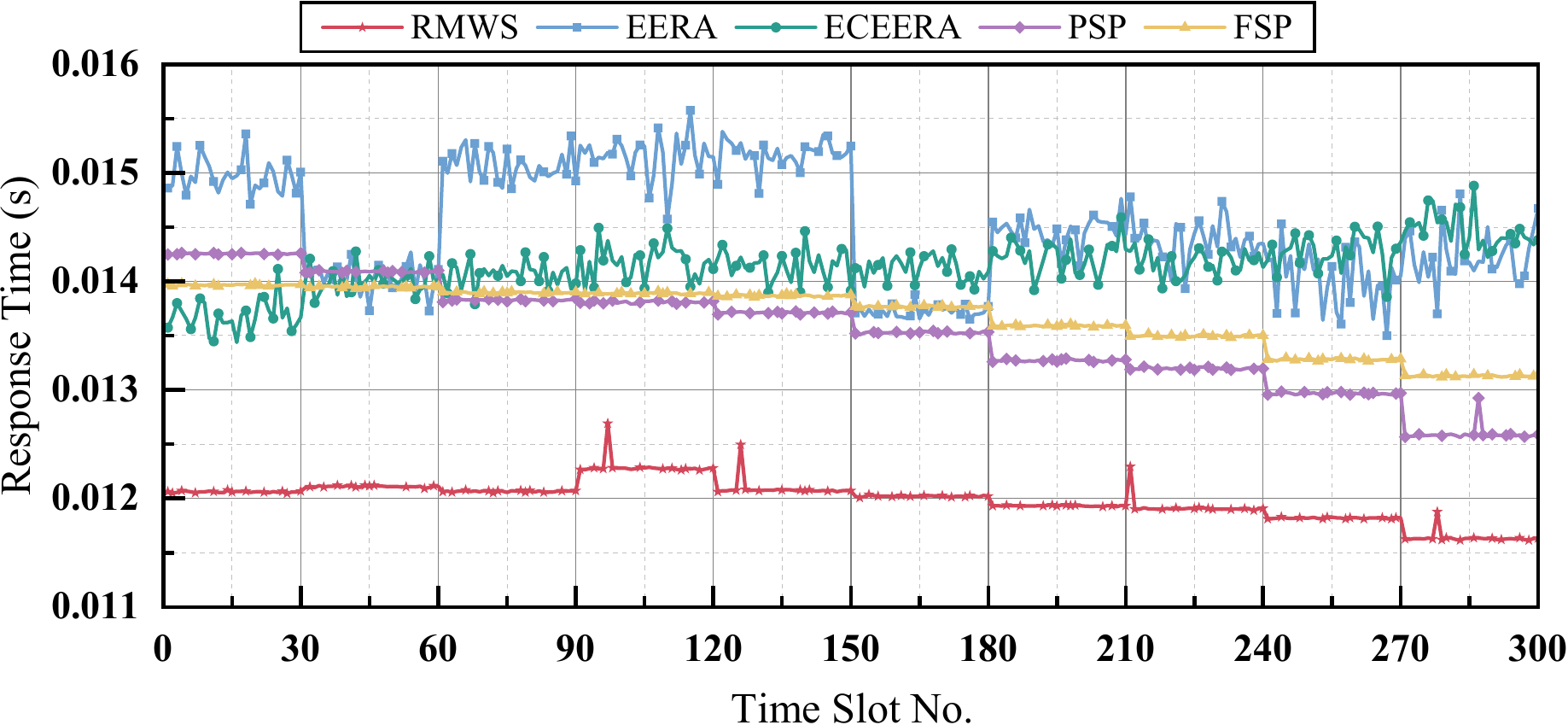}
        \label{frame_zipf_distribution_coefficient}
    }
    \  \ \
    \subfigure[The variations in the ranking of service popularity.]{
        \includegraphics[width=0.45\textwidth]{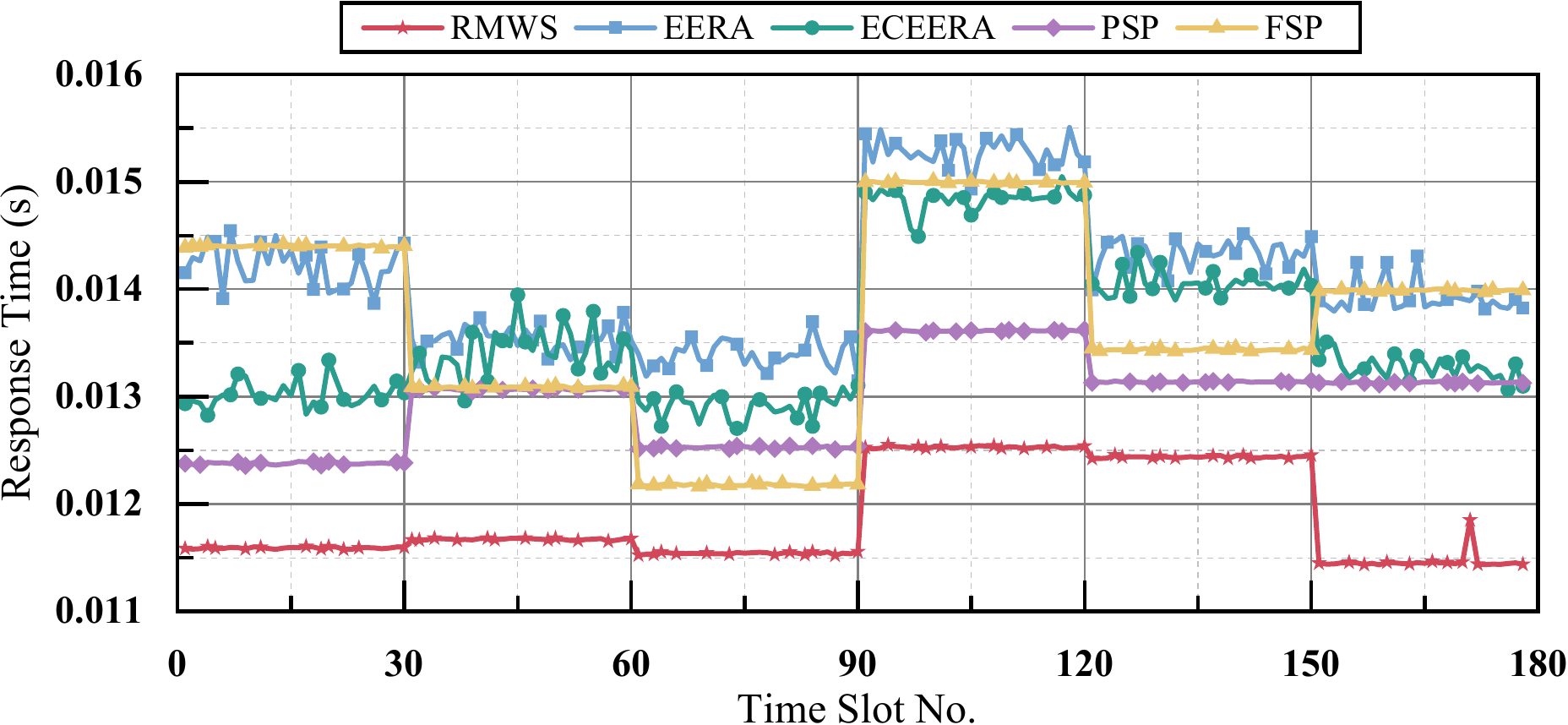}
        \label{frame_popular_rank}
    }
    % \hspace{0.1\textwidth} % 控制子图之间的水平间距
    \caption{Response time under different service popularity}
    \label{fig:service_popularity}
    \vspace{-10pt}
\end{figure*}

\item \textbf{Comparison under Different Storage
and Computation Capacity of Edge Servers.}
The capacity of computation and storage resources
in edge servers dictates the number
and distribution of services they can handle.
In~\Cref{edge_storage_capacity} and~\Cref{edge_computation_capacity},
% the storage and  computation resources of edge servers
% were increased from the default 40\% to 160\%.
It can be concluded that as the storage and  computation capacity of edge servers increase, the service response time of each algorithm gradually decreases. 
Notably, the RMWS consistently maintains
lower response delays
by effectively balancing service popularity and resource availability within the edge-cloud cooperative system.
Moreover,
in the absence of coordination between edge and cloud,
if edge servers have limited computation capabilities,
processing the entire workload on edge servers
can result in response delays surpassing those
incurred by offloaded to cloud servers.
Specifically, when the computation resources of the edge server are set to the default value of 40\%, the average response latency relationship of each algorithm is EERA $>$ CPO $>$ NSP, FSP, PSP, ECEERA, RWMS. This indicates that in cases where edge resources are insufficient, the edge-cloud collaboration mechanism can effectively reduce task processing latency. When the computation and storage resources of the edge server are set to the default value of 160\%, the performance of the non-collaborative service placement algorithm NSP is only better than CPO. This suggests that when edge resources are sufficient, cooperation among edge servers can also effectively reduce task processing time.

\item \textbf{Comparison under Different Storage
and Computation Requirement of Services.}
It can be seen from the~\Cref{service_storage_requirement}
and~\Cref{service_computation_requirement},
as the storage and computation demands of services increase, the number of services that edge servers can simultaneously deploy and tasks they can process both decrease. This will lead to an increase in service response time, but in such cases, RMWS consistently delivers optimal performance.
The RMWS demonstrates
notable efficiency by reducing service response latency
by  at least 10.6\%, 11.9\%, and 14.5\%
compared to other algorithms
when services require 60\%, 100\%, and 140\% 
of the default storage resources, respectively.
Similarly,
when services demand 60\%, 100\%, and 140\%
of the default computation resources,
our algorithm achieves a reduction of  at least 12.5\%, 11.7\%, and 7.2\% in  response latency
compared to other algorithms, respectively. This indicates that the RMWS has more performance in resource allocation and utilization.

% \begin{figure}[tbp]
% \centerline{\includegraphics[scale=0.3]{pic/Service_Storage_Requirement .pdf}}
% \caption{Response time under different storage requirement of services.}
% \label{service_storage_requirement}
% \end{figure}
% \begin{figure}[tbp]
% \centerline{\includegraphics[scale=0.3]{pic/Service_Computation_Requirement .pdf}}
% \caption{Response time under different computation requirement of services.}
% \label{service_computation_requirement}
% \end{figure}

 \item \textbf{Comparison under Different Service Popularity.}
% According to the experimental parameter settings, the algorithm will perform service placement and resource allocation every 30 minutes, and workload scheduling every minute. 
To validate the impact of service popularity on response time, we ensure that the total number of workload requests remains constant within each time frame, and adjust the popularity of services by modifying the coefficients of the Zipf distribution and the ranking of service popularity.
In~\Cref{frame_zipf_distribution_coefficient}, the experiment modifies the Zipf coefficient of service requests and randomly generates the Zipf distribution coefficients within 10 time frames: [0.23, 0.3, 0.41, 0.42, 0.46, 0.54, 0.64, 0.67, 0.76, 0.89].
In~\Cref{frame_popular_rank}, the experiment modifies the the ranking of service popularity and randomly generating 6 different rankings within 6 time frames.
It can be observed that without resource configuration optimization (EERA and ECEERA), the response latency of tasks in each time slot fluctuates significantly. Other algorithms that have undergone resource configuration optimization can maintain a relatively stable state. Additionally, service response latency is somewhat affected by service popularity, but the RMWS algorithm overall remains in a better state.

%  \begin{figure}[tbp]
% \centerline{\includegraphics[scale=0.28]{pic/frame_zipf_distribution_coefficient.pdf}}
% \caption{Response time under different service popularity.}
% \label{frame_zipf_distribution_coefficient}
% \end{figure}
% \begin{figure}[tbp]
% \centerline{\includegraphics[scale=0.28]{pic/frame_popular_rank.pdf}}
% \caption{Response time under different service popularity.}
% \label{frame_popular_rank}
% \end{figure}

\end{inparaenum}

In conclusion, the RMWS can achieve better performance based on dynamically changing workloads and resource requirements in terms of different metrics.

\section{Conclusions}
This paper introduces a cloud-assisted edge
computing framework focusing on service placement,
resource management, and workloads scheduling.
Considering the complexity of this problem
and the distinct optimization cycles
for different sub-problems,
we propose a two-timescale
optimization algorithm to minimize
workload response time.
Extensive simulations demonstrate
the effectiveness and advantages of
our algorithm in minimising the response delay.
For future work, we will investigate more complex scenarios, considering the mobility of IoT devices in the workload scheduling process.

\section{ACKNOWLEDGMENTS}
This work is supported by National Key R \& D Program of China (No. 2021YFB3300200), the National Natural Science Foundation of China (No. 62072451, 62102408, 92267105), Guangdong Basic and Applied Basic Research Foundation (No. 2024A1515010251, 2023B1515130002) and Guangdong Special Support Plan (No. 2021TQ06X990). %Shenzhen Basic Research Program (No. JCYJ20220818101610023), Shenzhen Industrial Application Projects of undertaking the National key R \& D Program of China (No. CJGJZD20210408091600002).
\appendix
\section{Proof}
\subsection{Proof of Theorem 1}
Let $A = \{a_{1},a_{2},...,a_{M}\}$ be the service placement decision space of all edge servers. At each iteration round, we randomly choose the edge server $k$ and its service placement decision from $A$. Following the iterations of algorithm, the service placement decisions $\mathcal{X}$ evolves as a $L$-dimension Markov chain in which the $i$th dimension corresponds to the $i$th edge server service placement decision. For the convenience of proving, let $L = 2$, so it is a two-dimensional Markov chain and denoted as $S_{x_{1},x_{2}}$. In each iteration, we change the service placement decision to $S_{x_{1}^{*},x_{2}}$ with the following probability:
\begin{flalign}
    \Pr(S_{x_{1}^{*},x_{2}}|S_{x_{1},x_{2}}) &= \frac{1}{2M} \times \frac{1}{1+e^{(\vartheta(S_{x_{1}^{*},x_{2}})-\vartheta(S_{x_{1},x_{2}}))/\omega}} \\
    &= \frac{1}{2M} \times \frac{e^{-\vartheta(S_{x_{1}^{*},x_{2}})/\omega}}{e^{-\vartheta(S_{x_{1}^{*},x_{2}})/\omega}+e^{-\vartheta(S_{x_{1},x_{2}})/\omega}} \notag.
\end{flalign}

Let the stationary distribution be $\Pr^{*}$, according to the detailed balance condition,
\begin{flalign}
    & \Pr{^{*}}(S_{a_{1},a_{1}})\Pr(S_{a_{1},a_{m}}|S_{a_{1},a_{1}}) \\
     & = \Pr{^{*}}(S_{a_{1},a_{m}})\Pr(S_{a_{1},a_{1}}|S_{a_{1},a_{m}}).  \notag
\end{flalign}
it can be derived that:
\begin{flalign}
   & \Pr{^{*}}(S_{a_{1},a_{1}}) \times \frac{1}{2M} \times \frac{e^{-\vartheta(S_{a_{1},a_{m}})/\omega}}{e^{-\vartheta(S_{a_{1},a_{m}})/\omega}+e^{-\vartheta(S_{a_{1},a_{1}})/\omega}} \label{con:1} \\
   & = \Pr{^{*}}(S_{a_{1},a_{m}})\times \frac{1}{2M} \times  \frac{e^{-\vartheta(S_{a_{1},a_{1}})/\omega}} {e^{-\vartheta(S_{a_{1},a_{m}})/\omega}+e^{-\vartheta(S_{a_{1},a_{1}})/\omega}} \notag.
\end{flalign}

By observation the above equation, we can find Eq. (\ref{con:1}) is symmetric and can be balanced  if the stationary joint distribution $\Pr{^{*}}(\Tilde{S}) = \gamma e^{-\vartheta(\Tilde{S})/\omega}$ for arbitrary state $\Tilde{S}$ in the strategy space $\Omega$, where $\gamma$ is a constant. Therefore, 
based on the probability conservation law, the stationary  distribution can be expressed as
\begin{flalign}
    \Pr{^{*}}(S_{x_{1},x_{2}}) = \frac{e^{-\vartheta(S_{x_{1},x_{2}})/\omega}}{\sum\limits_{S_{\Tilde{x}_{1},\Tilde{x}_{2}}\in \Omega} e^{-\vartheta(S_{\Tilde{x}_{1},\Tilde{x}_{2}})/\omega}} .\label{con:2}
\end{flalign}

Let $S_{x_{1}^{*},x_{2}^{*}}$ be the globally optimal service placement decisions, thus $\vartheta(S_{x_{1}^{*},x_{2}^{*}}) \leq \vartheta(S_{\Tilde{x}_{1},\Tilde{x}_{2}})$ has always been established for any $S_{\Tilde{x}_{1},\Tilde{x}_{2}}\in \Omega$. From Eq. (\ref{con:2}), we observe that $ \Pr{^{*}}(S_{x_{1}^{*},x_{2}^{*}})$ will increase with the decrease of $\omega$ and $\lim_{\omega \rightarrow 0}\Pr{^*}(S_{x_{1}^{*},x_{2}^{*}}) = 1$, which proves that the gibbs sample algorithm will converges to the optimal state in probability.

The above analysis can be directly extended to the $L$-dimensional Markov chain.
\subsection{Proof of Theorem 2}
According to the convex optimization theory, If the Hessian is a positive definite matrix, the optimization objective function is a convex function\cite{2004Convex}. By analysis, the optimization objective of problem \textbf{P2} can be transformed into:
\begin{flalign}
    & f(\mathcal{X}^*,\mathcal{Y},\mathcal{Z}) =\sum_{s=1}^{S} \bigg[ (1-\sum_{i=1}^{L}z_{i,s})n_{s}\phi_{c,s}\bigg] +\\& \sum_{i=1}^{L}\sum_{s \in \theta_{i}}\bigg[\frac{z_{i,s}n_{s}}{y_{i,s}F_{i}/c_{s}-z_{i,s}n_{s}/\Delta t}+  max\{z_{i,s}n_{s}-n_{i,s}, 0\}\cdot\phi_{s} \bigg]   .\notag
\end{flalign}

The Hessian of $f(y_{i,s},z_{i,s})$ is
\begin{align}
    \textbf{H}=\begin{bmatrix} 
H_{11}&H_{12}\\
H_{21}&H_{22}
\end{bmatrix}=\begin{bmatrix}
\dfrac{\partial^{2}f(y_{i,s},z_{i,s})}{\partial^{2}y_{i,s}}&\dfrac{\partial^{2}f(y_{i,s},z_{i,s})}{\partial y_{i,s}\partial z_{i,s}}\\[15pt]
\dfrac{\partial^{2}f(y_{i,s},z_{i,s})}{\partial z_{i,s}\partial y_{i,s}}&\dfrac{\partial^{2}f(y_{i,s},z_{i,s})}{\partial^{2} z_{i,s}}
\end{bmatrix}, \label{con:3}
\end{align}
\begin{flalign}
   &\bigtriangleup_{1}= H_{11}=\frac{2z_{i,s}n_{s}c_{s}F_{i}^{2}\Delta t^3}{(y_{i,s}F_{i}\Delta t-z_{i,s}n_{s}c_{s})^{3}}>0, & \label{con:4}
\end{flalign}
\begin{flalign}
    & \bigtriangleup_{2} =  \frac{-n_{s}^2c_{s}^2F_{i}^2\Delta t^4(y_{i,s}F_{i}\Delta t-z_{i,s}n_{s}c_{s})^{2}}{(y_{i,s}F_{i}\Delta t-z_{i,s}n_{s}c_{s})^{6}} < 0  \label{con:5}.  &
\end{flalign}
where $\bigtriangleup_{1}= H_{11}$ and $\bigtriangleup_{2} = H_{11}H_{22}-H_{12}H_{21}$. 

We can prove that the Hessian in Eq.(\ref{con:3}) is indefinite by proving that the leading principal minors of \textbf{H} have both positive and negative,which are given by Eq. (\ref{con:4}) and Eq. (\ref{con:5}). Therefore, the problem \textbf{P2} is a non-convex optimization problem.

\subsection{Proof of Theorem 3}
The optimization objective of problem \textbf{P3} can be denoted as $f(\mathcal{X}^*,\mathcal{Z}^*,\mathcal{Y})$, then the second-order derivative
of $f(\mathcal{X}^*,\mathcal{Z}^*,\mathcal{Y})$ is:
\begin{flalign}
    \frac{\partial^{2}f(\mathcal{X}^*,\mathcal{Z}^*,\mathcal{Y})}{\partial y_{i,s}\partial y_{j,s}} = 
    \begin{cases}
   \frac{2z_{i,s}^*n_{s}c_{s}F_{i}^{2}\Delta t^3}{(y_{i,s}F_{i}\Delta t-z_{i,s}^*n_{s}c_{s})^{3}} \geq 0 &\text{if } i=j \\
    0 &\text{else}
    \end{cases}. \label{con:6}
\end{flalign}

The parameters in Eq.(\ref{con:6}) are all positive, so the Hessian matrix of the objective function $f(\mathcal{X}^*,\mathcal{Z}^*,\mathcal{Y})$ is positive definite. And problem \textbf{P3} can be defined as a convex optimization problem since its constraints are linear.

The optimization objective of problem \textbf{P4} can be denoted as $f(\mathcal{X}^*,\mathcal{Y}^*,\mathcal{Z})$, then the second-order derivative
of $f(\mathcal{X}^*,\mathcal{Y}^*,\mathcal{Z})$ is:
\begin{flalign}
    \frac{\partial^{2}f(\mathcal{X}^*,\mathcal{Y}^*,\mathcal{Z})}{\partial z_{i,s}\partial z_{j,s}} = 
    \begin{cases}
   \frac{2y_{i,s}^*n_{s}^2c_{s}^2F_{i}\Delta t^2}{(y_{i,s}^*F_{i}\Delta t-z_{i,s}n_{s}c_{s})^{3}}  \geq 0 &\text{if } i=j \\
    0 &\text{else}
    \end{cases}. \label{con:7}
\end{flalign}

The parameters in Eq.(\ref{con:7}) are all positive, so the Hessian matrix of the objective function $f(\mathcal{X}^*,\mathcal{Y}^*,\mathcal{Z})$ is positive definite. And problem \textbf{P4} can be defined as a convex optimization problem since its constraints are linear.

% The optimization objective of problem \textbf{P4} can be denoted as $f(\mathcal{X}^*,\mathcal{Y}^*,\mathcal{Z})$, then the second-order derivative
% of $f(\mathcal{X}^*,\mathcal{Y}^*,\mathcal{Z})$ is:
% \begin{flalign}
%     \frac{\partial^{2}f(\mathcal{X}^*,\mathcal{Y}^*,\mathcal{Z})}{\partial z_{i,s}\partial z_{j,s}} = 
%     \begin{cases}
%    \frac{2y_{i,s}^*n_{s}^2c_{s}^2F_{i}\Delta t^2}{(y_{i,s}^*F_{i}\Delta t-z_{i,s}^*n_{s}c_{s})^{3}}  &\text{if } i==j \\
%     0 &\text{otherwise }
%     \end{cases}. \label{con:7}
% \end{flalign}

% The parameters in Eq.(\ref{con:7}) are all positive, so the Hessian matrix of the objective function $f(\mathcal{X}^*,\mathcal{Y}^*,\mathcal{Z})$ is positive definite. And problem \textbf{P4} can be defined as a convex optimization problem since its constraints are linear.

\subsection{Proof of Theorem 4}
By using the KKT conditions, we can obtain that:

\begin{equation}
\left\{
\begin{aligned}
&\frac{\partial L(f(\mathcal{Y}),\lambda,\mu)}{\partial y_{i,s}}=0 \\
&\lambda_{i}(\sum_{s\in \theta_{i}}y_{i,s}-1)=0 \\
&\mu_{i}\bigg[ \sum_{s\in \theta_{i}}\bigg(\frac{m_{s}}{M_{i}}P_{i}^{m}+y_{i,s}P_{i}^{f}\bigg)-P_{i}^{bud}\bigg]=0 \\
&\lambda_{i},\mu_{i} \geq 0
\end{aligned}
\right.
\end{equation}

Based on the above formulas, we can derive that:
\begin{equation}
\left\{
\begin{aligned} 
   & y_{i,s}^*=\frac{\sqrt{\frac{z_{i,s}^{*}n_{s}c_{s}F_{i}\Delta t^2}{\lambda_{i}+\mu_{i}P_{i}^{f}}}+z_{i,s}^{*}n_{s}c_{s}}{F_{i}\Delta t} \\
  & \sum_{s\in \theta_{i}}y_{i,s}\leq 1 \quad and \quad \sum_{s\in \theta_{i}}y_{i,s}\leq \Gamma_{i}  \quad  
  &
 \end{aligned}
\right.
\end{equation}

where $\Gamma_{i}=\bigg(p_{i}^{bud}-\sum_{s\in \theta_{i}}\frac{m_{s}}{M_{i}}P_{i}^{m}\bigg)/P_{i}^{f}$.

\textit{Case 1}: if $\Gamma_{i} < 1$, it means that $\sum_{s\in \theta_{i}}y_{i,s}\leq \Gamma_{i}$ and $\lambda_{i} = 0$.

\begin{flalign}
  &  \mu_{i}=\frac{(\sum_{s\in\theta_{i}}\sqrt{z_{i,s}^*n_{s}c_{s}F_{i}\Delta t^2})^2}{(\Gamma_{i}F_{i}\Delta t-\sum_{s\in\theta_{i}}z_{i,s}^*n_{s}c_{s})^2P_{i}^{f}}, &
\end{flalign}

\begin{flalign}
  & y_{i,s}^{*}=\frac{\sqrt{z_{i,s}^*n_{s}c_{s}}(\Gamma_{i}F_{i}\Delta t-\sum_{s\in\theta_{i}}z_{i,s}^*n_{s}c_{s})}{\sum_{s\in\theta_{i}}\sqrt{z_{i,s}^*n_{s}c_{s}}F_{i}\Delta t}+\frac{z_{i,s}^*n_{s}c_{s}}{F_{i}\Delta t}. &
\end{flalign}

\textit{Case 2}: if $\Gamma_{i} \geq 1$, it means that $\sum_{s\in \theta_{i}}y_{i,s}\leq 1$ and $\mu_{i} = 0$.

\begin{flalign}  &\lambda_{i}=\bigg(\frac{\sum_{s\in\theta_{i}}\sqrt{z_{i,s}^*n_{s}c_{s}F_{i}\Delta t^2}}{F_{i}\Delta t-\sum_{s\in\theta_{i}}z_{i,s}^*n_{s}c_{s}}\bigg)^2, &
\end{flalign}

\begin{flalign}
    &y_{i,s}^{*}=\frac{\sqrt{z_{i,s}^*n_{s}c_{s}}(F_{i}\Delta t-\sum_{s\in\theta_{i}}z_{i,s}^*n_{s}c_{s})}{\sum_{s\in\theta_{i}}\sqrt{z_{i,s}^*n_{s}c_{s}}F_{i}\Delta t}+\frac{z_{i,s}^*n_{s}c_{s}}{F_{i}\Delta t}.&
\end{flalign}

This ends the proof.
\cleardoublepage
\bibliographystyle{IEEEtran}
\bibliography{ref}
\end{document}